\documentclass[useAMS,usenatbib]{mn2e}
\usepackage{amsmath,lscape,graphicx,amssymb,times,multirow,epsfig,longtable,color,rotating,placeins}
\usepackage{natbib}

\title[The multiplicity of A-type stars within 75 pc]{The VAST
  Survey -- III. The multiplicity of A-type stars within 75 pc}

\author[De Rosa et al.]
{R. J. De Rosa$^{1,2}$\thanks{E-mail: rjderosa@asu.edu}\thanks{Based
    on observations obtained at the Canada-France-Hawaii Telescope
    (CFHT) under programme IDs 2008AC22, 2009BC06, 2010AC14 and
2011AC11. Based on observations obtained at the Gemini Observatory
under programme IDs  GN2008A-Q-74, GN-2008B-Q-119, GN-2010A-Q-75.
See Table \ref{tab:archive} for details of data obtained from the
CFHT and European Southern Observatory (ESO) archives.}, J. Patience$^{1,2}$,
  P. A. Wilson$^2$, A. Schneider$^3$, S. J. Wiktorowicz$^{4,5}$, \newauthor A. Vigan$^{6,2}$,
  C. Marois$^7$, I. Song$^3$,  B. Macintosh$^8$,
  J. R. Graham$^{4,9}$, R. Doyon$^{10}$, \newauthor
  M. S. Bessell$^{11}$, S. Thomas$^{12,13}$, \& O. Lai$^{14}$\\
$^1$ School of Earth and Space Exploration, Arizona State University,
PO Box 871404, Tempe, AZ 85287-1404, USA\\
$^2$ School of Physics, College of Engineering, Mathematics and Physical Sciences, University of Exeter, Stocker Road, Exeter, EX4 4QL,
UK\\
$^3$ Physics and Astronomy, University of Georgia, 240 Physics, Athens, GA 30602,
USA\\
$^4$ Department of Astronomy, University of California at Berkeley,
Berkeley, CA 94720, USA\\
$^5$ Department of Astronomy, University of California at Santa Cruz,
1156 High Street, Santa Cruz, CA 95064, USA\\
$^6$ Aix Marseille Universit\'e, CNRS, LAM (Laboratoire
d'Astrophysique de Marseille) UMR 7326, 13388, Marseille, France\\
$^7$ NRC Herzberg Institute of Astrophysics, 5071 West Saanich Road,
Victoria,
BC, V9E 2E7, Canada\\
$^8$ Institute of Geophysics and Planetary Physics, Lawrence Livermore
National Laboratory, 7000 East Ave, Livermore, CA 94550,
USA\\
$^{9}$ Dunlap Institute for Astronomy and Astrophysics, University of
Toronto, 50 St. George Street, Toronto, ON, M55 3H8, Canada\\
$^{10}$ D\'{e}partement de Physique, Universit\'{e} de Montr\'{e}al, C.P.
6128, Succ. Centre-Ville, Montr\'{e}al, QC,
H3C 3J7, Canada\\
$^{11}$ Research School of Astronomy and Astrophysics, Mount Stromlo Observatory, The Australian National University, ACT 2611,
Australia\\
$^{12}$ Gemini Observatory, Southern Operations Center, Casilla 603, La
Serena, Chile\\
$^{13}$ Laboratory for Adaptive Optics, University of California/Lick
Observatory,
University of California at Santa Cruz, 1156 High Street, Santa Cruz,
CA 95064, USA\\
$^{14}$ Canada-France-Hawaii Telescope, 65-1238 Mamalahoa Highway,
Kamuela, HI 96745, USA}
\date{-}
\pagerange{\pageref{firstpage}--\pageref{lastpage}} \pubyear{2013}
\begin{document}
\label{firstpage}

\maketitle

\begin{abstract} With a combination of adaptive optics imaging and a
multi-epoch common proper motion search, we have conducted a large
volume-limited ($D \le 75$ pc) multiplicity survey of A-type stars,
sensitive to companions beyond 30 au. The sample for the Volume-
limited A-STar (VAST) survey consists of 435 A-type stars: 363 stars
were observed with adaptive optics, 228 stars were searched for wide
common proper motion companions and 156 stars were measured with both
techniques. The projected separation coverage of the VAST survey
extends from 30 to 45,000 au. A total of 137 stellar companions were
resolved, including 64 new detections from the VAST survey, and the
companion star fraction, projected separation distribution and mass
ratio distribution were measured. The separation distribution forms a
log-normal distribution similar to the solar-type binary distribution,
but with a peak shifted to a significantly wider value of
$387^{+132}_{-98}$ au. Integrating the fit to the distribution over
the 30 to 10,000 au observed range, the companion star fraction for
A-type stars is estimated as 33.8 $\pm$ 2.6 per cent. The mass ratio
distribution of closer ($<125$ au) binaries is distinct from that of
wider systems, with a flat distribution for close systems and a
distribution that tends towards smaller mass ratios for wider
binaries. Combining this result with previous spectroscopic surveys of
A-type stars gives an estimate of the total companion star fraction of
$68.9\pm7.0$ per cent. The most complete assessment of higher order
multiples was estimated from the 156-star subset of the VAST sample
with both adaptive optics and common proper motion measurements,
combined with a thorough literature search for companions, yielding a
lower limit on the frequency of single, binary, triple, quadruple and
quintuple A-type star systems of  $56.4_{-4.0}^{+3.8}$,
$32.1_{-3.5}^{+3.9}$, $9.0_{-1.8}^{+2.8}$, $1.9_{-0.6}^{+1.8}$ and
$0.6_{-0.2}^{+1.4}$ per cent, respectively. \end{abstract}


\section{Introduction} Binary stars represent the most common product
of the star formation process and a key environmental factor for
planet formation. Investigating the properties of binary systems, and
dependencies on age (e.g. \citealp{Ghez:1997cg, Duchene:1999vw,
Patience:2002ik}), environment (e.g. \citealp{Kohler:2006ei,
King:2012cn, Sana:2013hh}) and primary mass (e.g.
\citealp{Abt:1983ki, Lada:2006kp, Lafreniere:2008kv, Kraus:2011hv,
Sana:2011cd}) is therefore crucial to our understanding of star
formation. Theoretical binary formation models (e.g.
\citealp{Boss:1979jk, Adams:1989ca, Bonnell:1991gt, Bonnell:1994vx,
Clarke:1996tl, Bate:1997uf, Kratter:2009gs, Stamatellos:2011ev}) and
the results of large numerical simulations of star formation within
stellar clusters (e.g. \citealp{Sterzik:2003jh, Moeckel:2010gg,
Bate:2011hy, Krumholz:2012bc}) require large-scale surveys for
empirical comparison.

Previous volume-limited surveys utilizing a range of companion
detection techniques, collectively sensitive to all possible binary
orbits, have provided observational constraints on the frequency and
properties of the binary companions to nearby field FGKM stars and
field L and T brown dwarfs (e.g. \citealp{Duquennoy:1991wk,
Fischer:1992cj, Burgasser:2006hd, Reid:2008gr, Raghavan:2010gd}). For
stars more massive than F-type stars, there are no comparable
comprehensive surveys covering the full range of binary orbits. A
large number of O and B-type stars have been observed with speckle
interferometry (e.g. \citealp{Mason:1997gf, Hartkopf:1999dd,
Mason:2009fm}); however, the samples typically consist of stars at
large distances and the technique is limited in magnitude difference
sensitivity and separation range coverage. Greater sensitivity to
fainter companion has been achieved using adaptive optics (AO) surveys
(e.g. \citealp{Roberts:2007ee}); however, the sample was magnitude-
limited, introducing a potential selection bias. For field A-type
stars, spectroscopic work has identified systems with short orbital
periods (e.g. \citealp{Abt:1965fz, Abt:1985bs, Carrier:2002hs,
Carquillat:2003di}). More recently, A-type stars have been the subject
of deep AO imaging searches for extreme mass ratio planetary
companions (e.g. \citealp{Vigan:2012jm, Nielsen:2013jy}), with several
planetary systems discovered \citep{Kalas:2008cs, Marois:2008ei,
Lagrange:2009hq, Marois:2010gp, Carson:2013fw, Rameau:2013dr}. As the
planetary population around A-type stars is revealed, the properties
of A-type star binaries will serve as an essential comparison.

This paper is the third in a series on the properties of the Volume-
limited A-STar (VAST) survey of A-type stars within 75 pc. The survey
was designed to have companion mass ratio sensitivity limits ($M_2/M_1
\gtrsim 0.1$) comparable to previous volume-limited multiplicity
surveys and to cover projected separations extending from the peak of
the solar-type distribution ($\sim$40 au). In previous papers, subsets
of the VAST survey were investigated to study the unexpected X-ray
emission of A-type stars \citep{DeRosa:2011ci} and orbital motion of
known binaries \citep{DeRosa:2012gq}. In this paper, the comprehensive
binary statistics are presented. The full sample is defined in Section
2, followed by the data acquisition of both new AO and literature wide
field imaging in Section 3. Data analysis techniques applied to
identify and characterize the candidate companions and detection
limits in both types of images are detailed in Section 4, and the
overall survey completeness is quantified in Section 5. The survey
results, including the A-type star binary separation distribution,
mass ratio distribution and companion star fraction (CSF), are
reported in Section 6. Comparisons of the VAST results with previous
survey and theoretical models are made in the discussion, Section 7,
along with a combination of the VAST results and previously known
binary companions to investigate the higher order multiple systems.
Finally, Appendix 1 describes the procedures employed to estimate the
mass and age of the VAST sample members.

\section{Sample}
\label{sec:sample}
\begin{figure}
\resizebox{\hsize}{!}{{\includegraphics{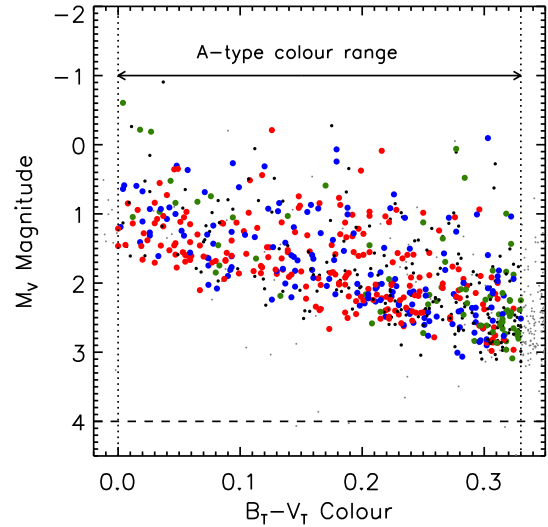}}} 
\caption{A colour--magnitude diagram demonstrating the selection
  criteria used to define the volume-limited sample of A-type
  stars. Of the {\textsl Hipparcos} stars within 75 pc (small
  grey filled points), those with a parallax uncertainty of
  $\sigma_{\pi}/\pi \le 0.05$ and with a $B_{\rm T}-V_{\rm T}$ colour consistent with A-type
  stars were selected (dotted vertical lines), with a magnitude
  cut-off to remove contamination from faint white dwarfs (dashed
  horizontal line). Of the 636 stars satisfying these
  criteria, 156 were observed with adaptive optics and
  are included within the photographic plate search for common proper
  motion companions (blue points), 207 targets were only observed with adaptive
  optics (red points), and 72 were only included within
  the photographic plate sample (green points). The remaining
  201 targets without observations, which otherwise satisfied the
  selection criteria, are plotted for reference (small black points).}
\label{fig:cmd}
\end{figure}
\begin{figure}
\resizebox{\hsize}{!}{{\includegraphics{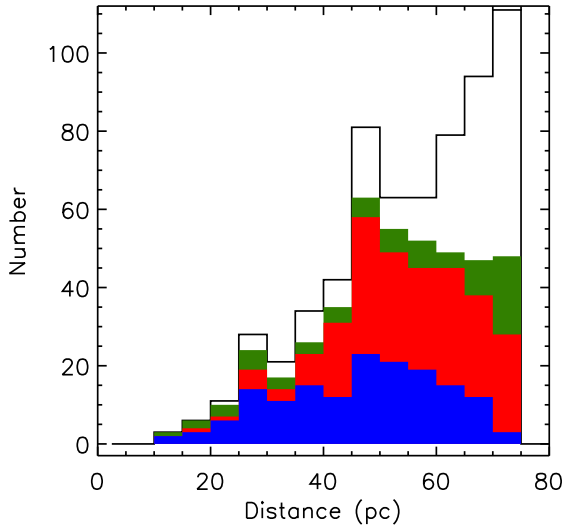}}} 
\caption{The distribution of distance estimates for the 636 A-type
  stars within the volume-limited sample. The shading of the histogram
  indicates whether the target was within both the adaptive optics and
  photographic plates sample (blue histogram), only the adaptive
  optics sample (red histogram), only the photographic plates
  sample (green histogram), or was not within either sample (open
  histogram). The distance distribution of the observed stars is
  complete up to approximately 50 pc, beyond which a significant
  proportion of the volume-limited sample remains
  unobserved. Targets at close distances were preferentially observed
  in order to increase the sensitivity to binary companions at small
  physical separations.}
\label{fig:distance}
\end{figure}
\begin{figure}
\resizebox{\hsize}{!}{{\includegraphics{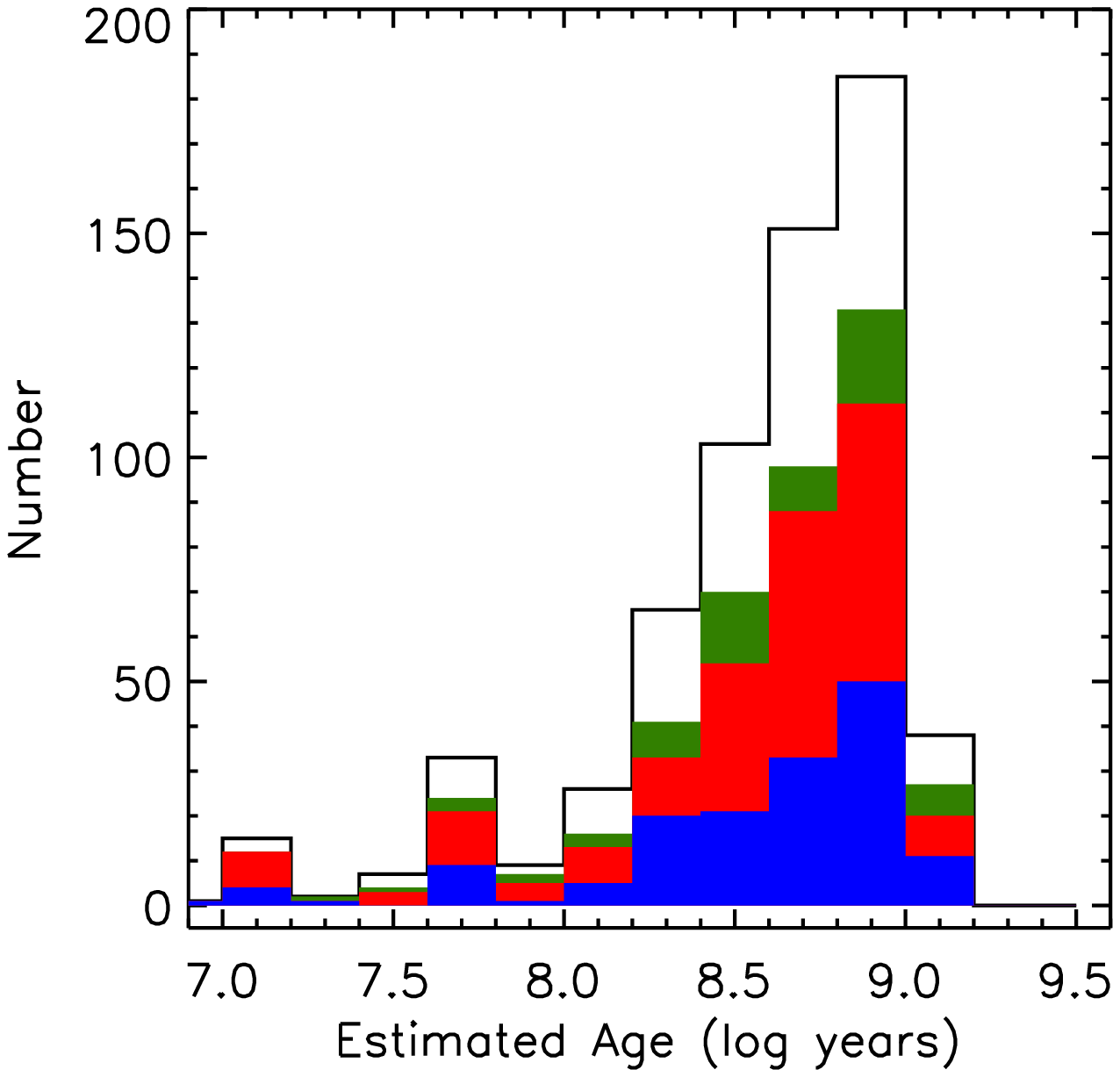}}} 
\caption{The distribution of the age estimates for each target estimated from
  solar-metallicity isochrones \citep{Siess:2000tk}. The shading of
  this histogram is as in Fig. 2.}
\label{fig:age}
\end{figure}
\begin{figure}
\resizebox{\hsize}{!}{{\includegraphics{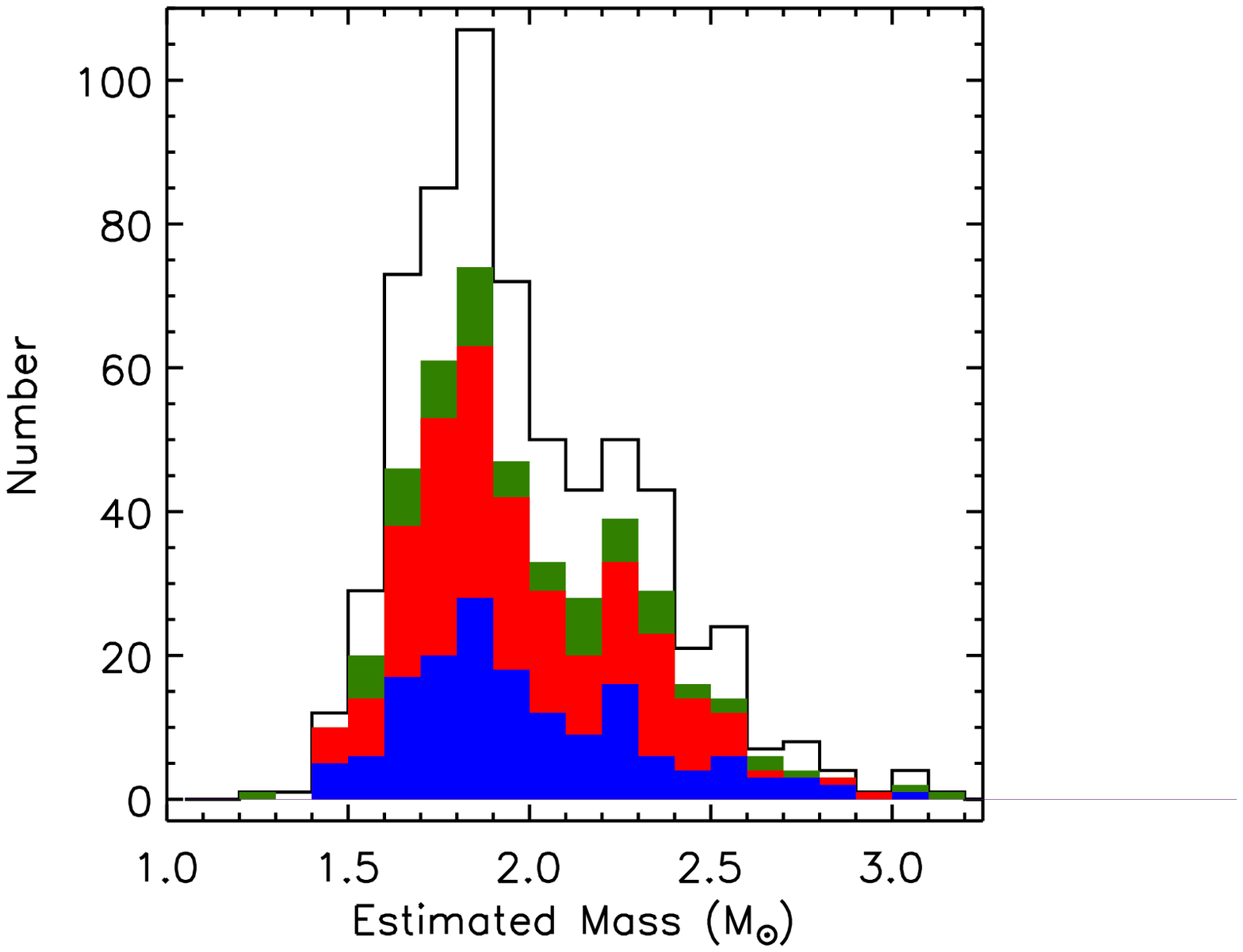}}} 
\caption{The distribution of the mass estimates for each target estimated from
  solar-metallicity isochrones \citep{Siess:2000tk}. The shading of
  this histogram is as in Fig. 2.}
\label{fig:mass}
\end{figure}

\begin{table*}
\caption{The VAST sample - available in its entirety in the electronic
edition of the journal.}
\begin{tabular}{ccr@{$\pm$}lr@{$\pm$}lr@{$\pm$}lr@{$\pm$}lr@{$\pm$}lr@{$\pm$}lcccr@{ $\to$ }lr@{ $\to$ }l}
HIP&Spectral&\multicolumn{2}{c}{Distance}&\multicolumn{2}{c}{$B_{\rm
    T}$}&\multicolumn{2}{c}{$V_{\rm
    T}$}&\multicolumn{2}{c}{$J$}&\multicolumn{2}{c}{$H$}&\multicolumn{2}{c}{$K_{\rm
    S}$}&Age$^{a}$&Age&Mass&\multicolumn{4}{c}{Separation
  Coverage ($\log a$ [au])}\\
&Type&\multicolumn{2}{c}{\textit{(pc)}}&\multicolumn{2}{c}{\textit{(mag)}}&\multicolumn{2}{c}{\textit{(mag)}}&\multicolumn{2}{c}{\textit{(mag)}}&\multicolumn{2}{c}{\textit{(mag)}}&\multicolumn{2}{c}{\textit{(mag)}}&\textit{(Myr)}&Ref.&\textit{($M_{\odot}$)}&\multicolumn{2}{c}{Adaptive
  Optics}&\multicolumn{2}{c}{CPM Search}\\
\hline
\hline       
128&Am...&70.8&1.7&6.73&0.01&6.52&0.01&6.10&0.02&6.06&0.04&6.02&0.02&180&1&1.84&1.50&2.90&3.60&4.65\\
159&A3&62.5&2.1&7.26&0.01&6.96&0.01&6.33&0.02&6.27&0.02&6.21&0.02&60&1&1.59&1.50&2.90&\multicolumn{2}{c}{--}\\
1473&A2V&41.3&0.4&4.59&0.01&4.51&0.01&4.34&0.27&4.42&0.18&4.46&0.29&200&1&2.26&1.50&2.55&3.60&4.65\\
2355&A7III&62.8&1.7&5.52&0.01&5.25&0.01&4.86&0.25&4.69&0.19&4.46&0.03&710&1&2.20&1.50&2.90&3.60&4.65\\
2381&A3V&53.1&0.8&5.32&0.01&5.18&0.01&5.28&0.24&4.88&0.08&4.83&0.02&450&1&2.09&1.50&2.90&\multicolumn{2}{c}{--}\\
\hline
\multicolumn{21}{l}{$a$ - Age rounded to nearest isochrone value. Age estimates from: 1 - this work (CMD), 2 - \cite{Zuckerman:2011bo}, 3 - \cite{Tetzlaff:2010gt}, 4 - \cite{Rhee:2007ij},}\\
\multicolumn{21}{l}{5 - \cite{Su:2006ce}, 6 - \cite{Rieke:2005hv}, 7 - \cite{Perryman:1998vj}, 8 - \cite{BarradoyNavascues:1998tg}, 9 - \cite{Song:2001bv}, 10 - \cite{Westin:1985un},}\\
\multicolumn{21}{l}{11 - \cite{Gerbaldi:1999cv}, 12 - \cite{Zuckerman:2004ex}, 13 - \cite{Laureijs:2002de}, 14 - \cite{Stauffer:1995jl}, 15 - \cite{Paunzen:1997vj}, 16 - \cite{Janson:2011hu},}\\
\multicolumn{21}{l}{17 - \cite{Torres:2008vq}.}\\
\hline
\label{tab:sample}
\end{tabular}
\end{table*}
\begin{table*}
\caption{Alternative catalogue identifiers - available in its entirety in the electronic
edition of the journal.}
\begin{tabular}{cccccccc}
 HIP&Name&Bayer&Flamsteed&HR&HD&ADS&WDS\\
 \hline
 \hline
 128&&&&&224890&&\\
 159&&&&&224945&&\\
 1473&&$\sigma$ And&25 And&68&1404&&\\
 2355&&&28 And&114&2628&409 AB&J00301+2945\\
 2381&&&&118&2696&&\\
\hline
\label{tab:alternative}
\end{tabular}
\end{table*}

To measure the frequency of stellar binary companions, and the
distribution of their separations and mass ratios, we have obtained
observations of a sample of 435 nearby A-type stars. The sample is
composed of two overlapping sets of A-type stars within 75 pc: a 363
star sample observed with AO instrumentation and a  228 star sample
investigated with astrometry obtained from all-sky photographic
surveys, with an overlap of 156 stars. The 435 observed targets,
listed in Tables \ref{tab:sample} and \ref{tab:alternative}, were
drawn from a volume-limited sample of A-type stars selected from the
\textsl{Hipparcos} catalogue \citep{ESA:1997ws, vanLeeuwen:2007dc}.
The sample was limited to targets within 75 parsecs, corresponding to
an \textsl{Hipparcos} parallax of $\pi \ge 13.3$ mas. High quality
parallax uncertainties ($\sigma_{\pi}/\pi \le 0.05$) are required to
place the targets on the colour--magnitude diagram, determine accurate
distances to the targets and, consequently, determine the absolute
magnitude of any resolved companion candidate. Using the optical
magnitudes of each target from the Tycho2 catalogue
\citep{Hog:2000wk}, the sample was limited to targets within the
A-type star colour range ($0.0 \le B_{\rm T}-V_{\rm T} \le 0.33$;
\citealp{Gray:1992wb}). Finally, an absolute magnitudecut-off of $M_V
< 4$ was imposed to remove two faint white dwarfs that have colours
consistent with A-type stars. Due to their brightness exceeding the
magnitude limit of the Tycho2 catalogue ($V_{\rm T} \lesssim 2.1$;
\citealp{Hog:2000wk}), seven nearby A-type stars were not listed
within the catalogue - $\alpha$ CMa (Sirius), $\alpha$ Gem, $\beta$
Leo, $\alpha$ Oph, $\alpha$ Lyr (Vega), $\alpha$ Aql (Altair) and
$\alpha$ PsA (Fomalhaut) - and were therefore excluded by the sample
selection process.

Drawn from the 636 stars which satisfy these selection criteria, the
VAST sample consists of 435 A-type stars within 75 pc. The
positions of the VAST targets on the colour--magnitude diagram are
shown in Fig. \ref{fig:cmd}, and span the full range of A-type
stars. A histogram of the distances to the observed targets is shown
in Fig. \ref{fig:distance}, with the VAST sample nearly complete
within a distance of 50 pc. In order to convert the measured
magnitude difference between an A-type star primary and a resolved
companion into a mass ratio for the system, an estimate of the age is
required. This is due to the age dependence of the mass-magnitude
relation used within this study for A-type stars, and for low-mass
M-dwarfs at ages $\lesssim$100 Myr. The procedure used for estimating
the age and mass of each target is given in Appendix 1, with the
resulting distribution of ages and masses of the sample given in
Figs \ref{fig:age} and \ref{fig:mass}.

\section{Data Acquisition}
\subsection{Adaptive optics observations}
\begin{table*}
\caption{Observing run details}
\begin{tabular}{cccccc}
Telescope&Instrument&Programme ID&Narrow&Wide&Observed\\
&&&filter&filter&stars$^{a}$\\
\hline
\hline
CFHT&KIR&2008AC22&FeII&$H$&31\\
&&2008AC22&$H_2 \left(v=1-0\right)$&$K^{\prime}$&8\\
&&2009BC06&$H_2 \left(v=1-0\right)$&$K^{\prime}$&46\\
&&2010AC14&$H_2 \left(v=1-0\right)$&$K^{\prime}$&42\\
&&2011AC11&$H_2 \left(v=1-0\right)$&$K^{\prime}$&5\\
Gemini North&NIRI&GN-2008A-Q-74&Br$\gamma$&$K^{\prime}$&36\\
&&GN-2008B-Q-119&Br$\gamma$&$K^{\prime}$&78\\
&&GN-2009B-Q-120&Br$\gamma$&$K^{\prime}$&3\\
&&GN-2010A-Q-75&Br$\gamma$&$K^{\prime}$&39\\
Lick&IRCAL&--&Br$\gamma$&$K_{\rm S}$&81\\
&&2012 SO16&$H_2 \left(v=1-0\right)$&$K_{\rm S}$&13\\
Palomar&PHARO&--&CH4$_{\rm S}$&($H$)$^{b}$&34\\
&&--&Br$\gamma$&($K$)$^{b}$&31\\
&&--&$K_{\rm S}$&($K$)$^{b}$&8\\
\hline
\multicolumn{6}{l}{$^{a}$ - These totals include targets with
  multiple epochs of observations}\\
\multicolumn{6}{l}{$^{b}$ - The wide band filter was not used for
data obtained with PHARO}\\
\multicolumn{6}{l}{KIR -- \cite{Doyon:1998vi}}\\
\multicolumn{6}{l}{NIRI (Near InfraRed Imager and Spectrometer) -- \cite{Hodapp:2003ko}}\\
\multicolumn{6}{l}{IRCAL (IR Camera for Adaptive Optics at Lick) -- \cite{Lloyd:2000we}} \\
\multicolumn{6}{l}{PHARO (Palomar High Angular Resolution Observer) --
  \cite{Hayward:2001kq}}\\
\label{tab:instruments}
\end{tabular}
\end{table*}
\begin{table}
\centering
\caption{Sources of archive observations}
\begin{tabular}{cc | cc}
\multicolumn{2}{c}{CFHT archive}&\multicolumn{2}{c}{ESO archive}\\
\hline
Programme ID&PI&Programme ID&PI\\
\hline
\hline
97IIH06&Simon&070.C-0565(A)&Mouillet\\
98IF12&Corcoran&272.D-5068(A)&Ivanov\\
98IH02&Simon&073.C-0469(A)&Chauvin\\
99IF59&Perrier&074.D-0180(A)&Ivanov\\
00BF1&Gerbaldi&076.C-0270(A)&Galland\\
01AF11&Gerbaldi&076.D-0108(A)&Ivanov\\
01AH11A&Jewitt&077.D-0147(A)&Ivanov\\
01BF21&Catala&079.C-0908(A)&Zuckerman\\
02AF03&Catala&080.D-0348(A)&Ivanov\\
03BH59A&Ftaclas&081.C-0653(A)&Lagrange\\
06BF07&Galland&382.D-0065(A)&Kervella\\
07BF04&Lagrange&383.C-0847(A)&Schmidt\\
08AF02/F07&Beuzit&&\\
08AF07&Lagrange&&\\
\hline
\end{tabular}
\label{tab:archive}
\end{table}

High-resolution AO images of a sample of 363 A-type stars were
obtained through a combination of new observations of 257 stars
obtained between 2008 and 2011 and archive images of an additional 106
targets. The details of the new AO observations are reported in Table
\ref{tab:instruments}, and the archive AO data included in the sample
are summarized in Table \ref{tab:archive}. For the new observations,
images were obtained using a near-infrared filter, decreasing the
contrast between the bright A-type star primary and any faint
companion candidate relative to optical wavelengths. The observational
strategy was designed such that the images would be sensitive to
companions at the bottom of the main sequence ($\Delta K \approx
7-10$, depending on the primary), at angular separations of $\rho \ge
1$ arcsec. Unsaturated exposures of each target were taken to detect
high mass ratio companions at close separations and calibrate the
photometry. Longer saturated exposures were also obtained and provided
sensitivity to low-mass companions beyond the saturated point spread
function of the bright target. A typical unsaturated sequence included
exposures obtained with a low transmission narrow-band filter at a
number of dither positions, removing the effect of bad pixels and
cosmic ray events. For the observations involving CFHT/KIR, Lick/IRCAL
and Gemini/NIRI, saturated exposures were then taken using a wide-band
filter at a number of dither positions, significantly increasing the
sensitivity of the observations to faint companions at the bottom of
the main sequence. Saturated exposures were not obtained using
Palomar/PHARO; instead, a large number of short exposures were
combined to achieve sensitivity to faint companions.

In addition to the new observations presented within this study,
VLT/NaCo \citep{Lenzen:2003iu,Rousset:2003hh} and CFHT/KIR data taken
in similar AO snapshot modes were obtained from the ESO and CFHT
science archives. Archived observations which were obtained using
complex imaging techniques such as angular differential imaging,
primarily used to search for extremely faint planetary-mass companions
\citep{Marois:2006df}, are not included in this study. AO images of
184 of the sample members were obtained from the two science archives,
providing measurements of an additional 106 targets not observed as a
part of our dedicated observing programme.

\subsection{Digitized photographic plates}
\begin{table}
\centering
\caption{Sources of photographic plates}
\begin{tabular}{cccc}
Survey&Filter&Declination range&Date range\\
\hline
\hline
ESO&$R$&$-90.0<\delta<-17.5$&1979--1990\\
POSS-I&$R$&$-20.5<\delta<+2.5$&1949--1957\\
POSS-I&$R$&$+2.5<\delta<+90.0$&1949--1957\\
POSS-II&$B$&$+2.5<\delta<+90.0$&1986--2002\\
POSS-II&$R$&$+2.5<\delta<+90.0$&1986--1999\\
POSS-II&$I$&$+2.5<\delta<+90.0$&1989--2000\\
UKST&$B$&$-90.0<\delta<+2.5$&1974--1993\\
UKST&$R$&$-90.0<\delta<+2.5$&1984--1999\\
UKST&$I$&$-90.0<\delta<+2.5$&1978--2002\\
\hline
\end{tabular}
\label{tab:plates}
\end{table}

The high angular resolution AO component of the survey was augmented
with a search of astrometric catalogues for wide common proper motion
(CPM) companions, confirmed with a visual inspection of digitized
photographic plates. Astrometric data were obtained from a number of
sources: bright CPM companions were identified from a previous
analysis of the {\sl Hipparcos} catalogue \citep{Shaya:2010dt}, while
fainter CPM companions were selected from a union of the SuperCosmos
Sky Survey (SSS) Science Archive \citep{Hambly:2001kf}, the PPMXL
catalogue \citep{Roeser:2010cr} and the Fourth US Naval Observatory
CCD Astrograph Catalog (UCAC4; \citealp{Zacharias:2012um}). In order
to estimate the innermost separation at which these catalogues were
sensitive to all stellar companions, and to select a subsample of
stars with a high apparent motion, digitized scans of photographic
plates from the UK Schmidt (UKST), ESO Schmidt and Palomar Oschin
Schmidt (POSS) sky surveys (Table \ref{tab:plates}) were obtained from
the SSS Science Archive \citep{Hambly:2001kf} for each of the 636
stars within the sample.

\section{Data reduction and candidate identification and characterization}
\subsection{Adaptive optics observations}

Each science image was processed through the standard near-infrared
data reduction process, beginning with a dark frame subtraction, and
division by a flat-field. The sky background for each target was
estimated from a median combination of the science images obtained at
different dither positions, unless dedicated sky frames were taken,
and was scaled to and subtracted from each science image. Bad pixels
and cosmic ray events were flagged and interpolated over, using both a
bad pixel map, generated from the flat-field analysis, and a search
for pixel outliers within each image. To increase sensitivity to faint
companions and increase the signal-to-noise ratio of any detection,
the individual science images obtained for each target were aligned to
a common centre and combined through a median combination.

The registration of the unsaturated images was achieved by determining
the Gaussian centroid of the target within each image; close
companions were masked before centroiding. The images of each target
were then shifted using cubic interpolation. Prior to performing a
median combination of the images, a radial profile was calculated and
subtracted. The median of these aligned images was then calculated to
create the final science image. For the saturated images, an estimate
of the centroid of the target was determined through an analysis of
the diffraction spikes from the secondary mirror supports on each
telescope. By cross-correlating the position of the diffraction spikes
within each individual saturated exposure of a given target, the
offsets required to register each image to a common centre were
calculated \citep{Lafreniere:2007cv}. As with the unsaturated
exposures, a radial profile was calculated and subtracted from each
aligned image, and a final science image was created through a median
combination.

Companion candidates were identified through a visual inspection of
the final science images. The centroid of each companion candidate was
compared with the centroid of the target in order to measure the
separation and position angle of the companion. These measurements
were converted into an on-sky separation and position angle
($\rho,\theta$) using the pixel scale and angle of North of the
detector, estimated from observing an astrometric field (the
Trapezium; \citealp{McCaughrean:1994id}) or a calibration binary. The
magnitude difference measured between each companion candidate and the
central target was estimated using aperture photometry, with the flux
of the central target scaled as a function of the filter transmission
and exposure time for companions resolved within the saturated
exposures. The aperture was twice the full width at half-maximum
(FWHM) measured within the final science frame, with a sky annulus
between six and eight times the FWHM. The uncertainties of the separation,
position angle and magnitude difference were estimated from the
standard deviation of the astrometric and photometric measurements
obtained from each individual exposure prior to combination.

The angular separation of each companion was converted to a projected
separation ($a_{\rm proj}$) using the distance to the primary obtained
from the \textsl{Hipparcos} catalogue \citep{vanLeeuwen:2007dc}. The
magnitude difference between each primary and companion was converted
to both a secondary mass ($M_2$) and mass ratio ($q=M_2/M_1$), using
the absolute magnitude of the primary obtained from 2MASS and
theoretical solar-metallicity isochrones
\citep{Siess:2000tk,Baraffe:1998ux}. For candidates at close
separations ($\lesssim 5$ arcsec), the absolute magnitude of the
primary obtained from the 2MASS catalogue is based on the sum of the
flux from both primary and companion. The contamination of the
magnitude of the primary caused by the presence of a bright companion
was removed using the magnitude difference measured within this study,
prior to the estimation of the mass of the companion and the mass
ratio of the system.

\begin{figure*}
\resizebox{\hsize}{!}{{\includegraphics{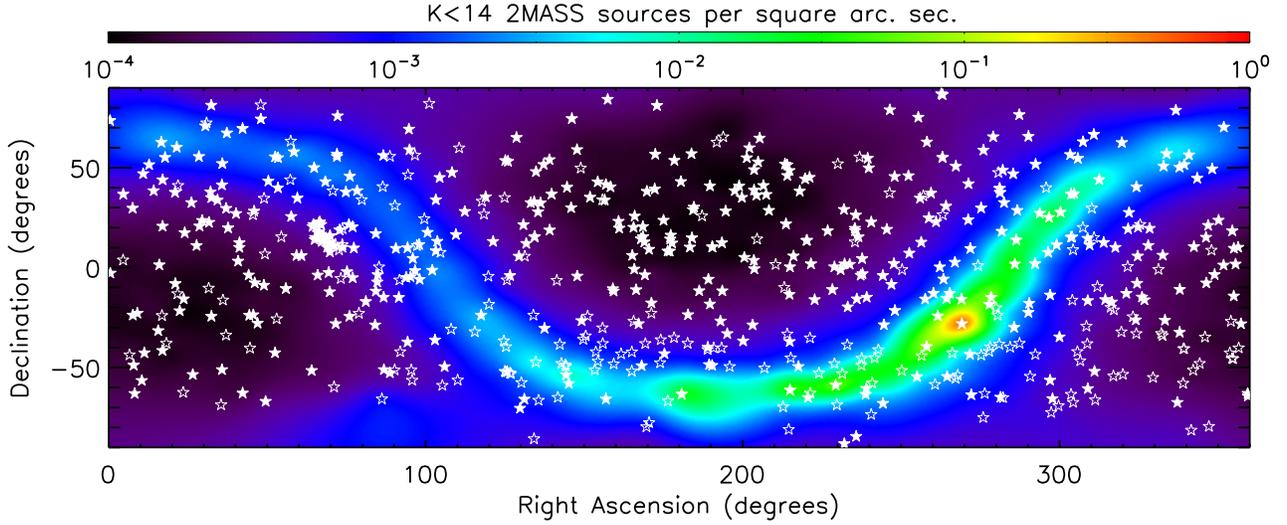}}} 
\caption{The sample target members are distributed evenly
  throughout the celestial sphere (open and filled star symbols), with
  the exception of the Hyades open cluster. The observed and
  unobserved VAST sample members are shown as filled and open stars,
  respectively. The surface density of 2MASS
    point sources with a magnitude of $K_{\rm S}<14$, within an area of
    1 square arc second surrounding each member of the VAST
    sample, is depicted by the coloured surface. The path of the
    Galaxy through the sky is apparent, with the Galactic Centre
    showing a significant increase in the surface density of
    2MASS point sources $N \left(K_{\rm S}<14\right) > 10^{-1}$.}
\label{fig:backgroundprob}
\end{figure*}

With only one epoch for most targets, the assessment of physical
association was based on a statistical cut-off estimated using 2MASS
source counts. The spatial distribution of the targets, superimposed
on the surface density of 2MASS point sources, is shown in Fig.
\ref{fig:backgroundprob}. Using the 2MASS catalogue, a power law was
fit to the cumulative number of 2MASS $J$, $H$ and $K_{\rm S}$ point
sources per square arcsecond as a function of magnitude
\citep{DeRosa:2011ci}. For each companion candidate resolved within
this study, the corresponding probability of finding a 2MASS point
source of the same magnitude or brighter within the same radius was
calculated. Only companion candidates with less than 5 per cent chance
of being a chance superposition were included in the final results and
discussion.

\subsection{Digitised photographic plates}
\begin{figure}
\resizebox{\hsize}{!}{{\includegraphics{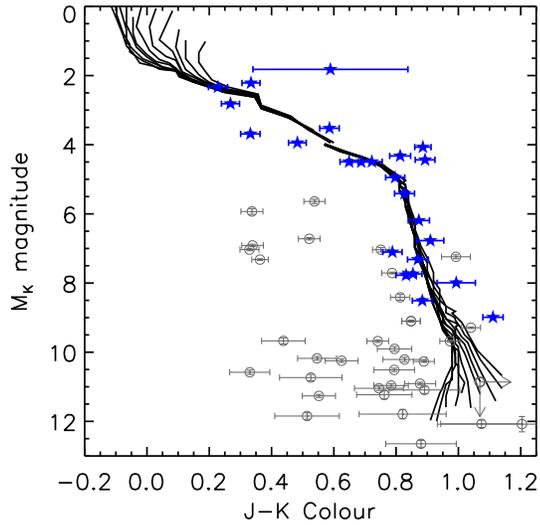}}} 
\caption{The position of each common proper motion candidate on the
  colour--magnitude diagram was compared with theoretical isochrones
  (black lines, \citealp{Baraffe:1998ux, Siess:2000tk}) in order to reject background
  objects with the same apparent motion. Those candidates thought to
  be physically associated based on their proper motion and position
  on the colour--magnitude diagram are highlighted (blue
  stars). Candidates with mass ratios of $q<0.05$, but otherwise
  consistent with being physically associated, were excluded.}
\label{fig:cpmcmd}
\end{figure}

In order to ensure that the CPM companion detections were reliable, a
subsample was designed to remove those targets with very small proper
motions. As the astrometric data obtained from the SSS Science Archive
represent the data set with the shortest time baseline over which to
measure proper motions, a lower limit on the total expected proper
motion of a bound CPM of 4 pixels ($>2.7$ arcsec) was applied to the
complete sample. Using the longest time baseline between the
photographic plates obtained from the SSS Science Archive, and the
annual proper motions reported within the {\sl Hipparcos} catalogue, a
sample of 228 stars was selected. As the level of sensitivity to
faint stellar companions varies as a function of separation, and the
brightness of target itself, the innermost separation to which the
astrometric catalogues were sensitive to all stellar companions was
estimated from the photographic plates. For each of the 228 targets,
the innermost separation was estimated as the separation at which the
flux of the primary reduced to 75 per cent of its peak value, a
conservative estimate based on an analysis of the magnitudes of the
background objects within each field.

Bright ($V\lesssim 8$) CPM companions were drawn from an analysis of
the {\sl Hipparcos} catalogue, taking into account both the proper
motions and parallax measurements \citep{Shaya:2010dt}. Companions
below the limiting magnitude of the {\sl Hipparcos} catalogue were
identified based on their proper motion from the three astrometric
catalogues described previously; the SSS Science Archive
\citep{Hambly:2001kf}, the PPMXL catalogue \citep{Roeser:2010cr} and
the UCAC4 catalogue \citep{Zacharias:2012um}. Only those objects with
a proper motion within $1.5\sigma$ of the motion of the primary stated
within the {\sl Hipparcos} catalogue were selected. In order to ensure
a CPM companion was not missed during this automated procedure, the
position of each source within the astrometric catalogues were marked
within the mosaic of each target, revealing the presence of any source
not included within the astrometric catalogues.

The relative position and brightness of each identified CPM companion
was determined from the astrometry and photometry within the 2MASS
catalogue, from which the projected separation and position angle
were calculated. The magnitude difference between the target and the
comoving companion was converted into a secondary mass and a mass
ratio using theoretical solar-metallicity isochrones
\citep{Baraffe:1998ux, Siess:2000tk}, in the same manner as the AO
companions. Finally, the CPM companions were plotted on a colour--
magnitude diagram (Figure \ref{fig:cpmcmd}), to reject background
objects with similar proper motions but an unphysical location on the
colour--magnitude diagram, assuming the primary and companion are at
the same distance.

\section{Survey Completeness}
\label{sec:comp}
\begin{figure}
\resizebox{\hsize}{!}{{\includegraphics{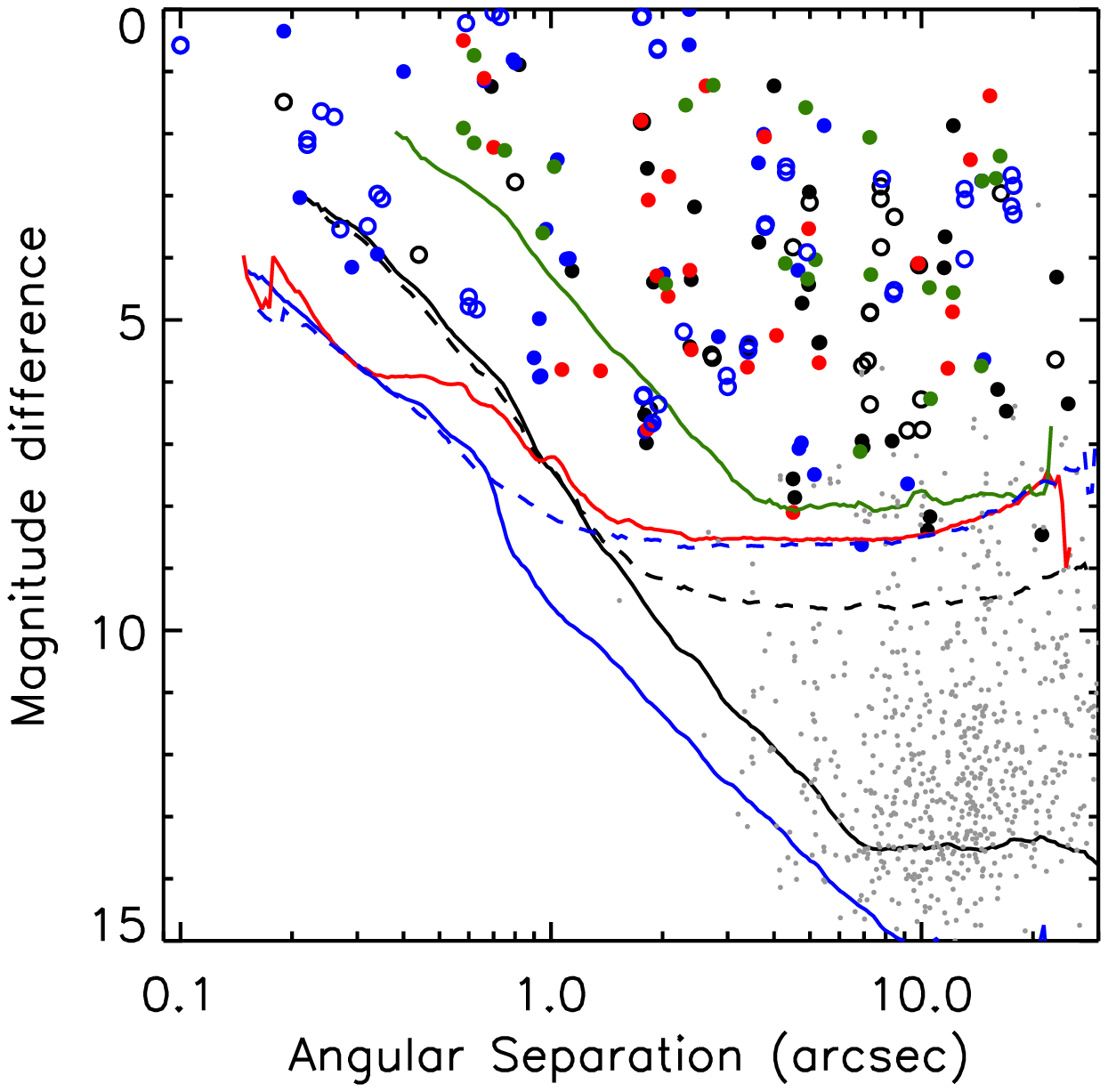}}} 
\caption{The average sensitivity of the observations obtained using
  KIR at the CFHT (black solid curve), NIRI at Gemini North (blue
  solid curve), PHARO at Palomar (red solid curve), IRCAL at the Lick
  Observatory (green solid curve), and observations obtained from the
  CFHT/KIR (black dashed curve), and ESO/NaCo (blue dashed curve) science
  archives. Companion candidates have been coloured corresponding to
  the instrument with which they were detected, with open circles
  corresponding to the two archive sources. Those companion candidates
  which fail the statistical criterion described in this section, or
  have mass ratios of $q<0.1$ are denoted by small grey
  points. Duplicate observations of the same companion have not been
  removed from this figure.}
\label{fig:tellimits}
\end{figure}
\begin{figure*}
\resizebox{\hsize}{!}{{\includegraphics{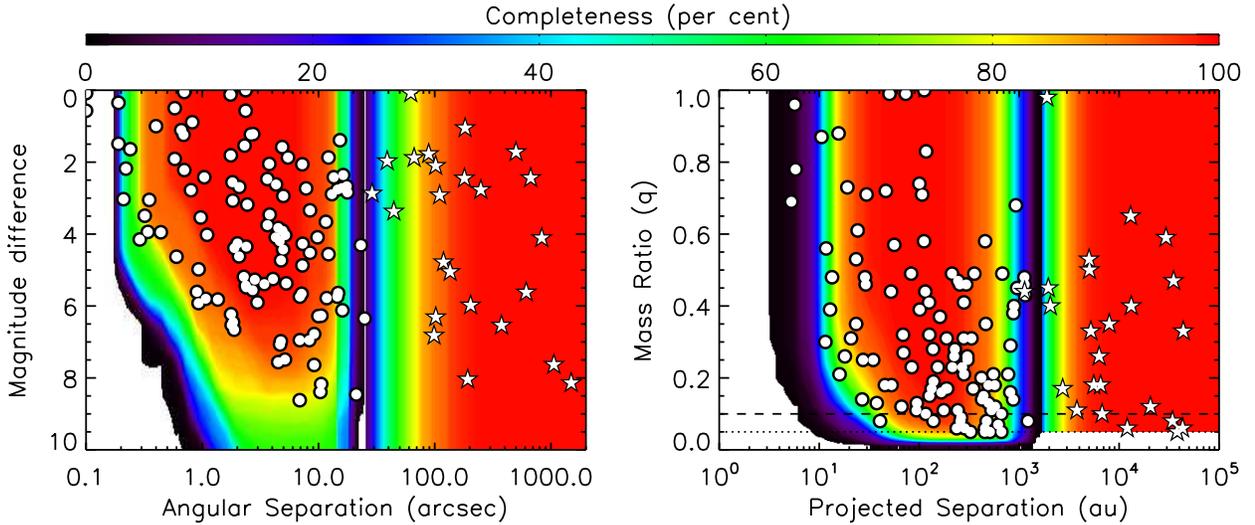}}} 
\caption{The completeness of the adaptive optics observations
  ($N=363$, $\rho \lesssim 15$ arcsec, $a_{\rm proj} \lesssim 2,000$ au) and the
  analysis of the photographic plates ($N=228$, $\rho \gtrsim
15$ arcsec, $a_{\rm proj} \gtrsim 2,000$ au), expressed in terms of
  (\textit{left-hand panel}) angular separation and magnitude difference,
  and (\textit{right-hand panel}) projected separation and mass
  ratio. The companion candidates identified within the adaptive
  optics data set (white filled points) and CPM companions identified
  within the photographic plates (white filled stars) are
  plotted. The regions of phase space not sampled by the observations
  are shown in white for clarity.}
\label{fig:completeness}
\end{figure*}

The completeness of the observations was estimated in order to
minimize any bias within the distribution of companion properties
measured within the survey. For the AO data, a one-dimensional
sensitivity curve was created for each image, with the sensitivity
limit at radius $r$ calculated as the standard deviation of the image
pixel values within a circular annulus of width $2\lambda/D$. The
faintest companion to which the data were sensitive was estimated as a
signal five times this standard deviation. The largest separation
considered for each target was the radius at which at least 90 per
cent of the position angles were sampled by the images. For targets
with multiple observations, the best contrast achieved at a given
separation was used as the formal detection limit at that separation.
The average detection limits for observations obtained using each
instrument are shown in Fig. \ref{fig:tellimits}, while the overall
completeness of the survey is given in terms of both observable
($\rho$, $\Delta$m) and physical quantities ($a_{\rm   proj}$, $q$) in
Fig. \ref{fig:completeness}.

\section{Results}
\onecolumn
\begin{longtable}{ccccccccccc}
\caption{Binary companions identified within the adaptive optics observations}\\
HIP&WDS&$\rho$&$a_{\rm proj}$&$\theta$&Band&$\Delta m$&$M_1$&$M_2$&$q$&Date\\
&Desig.&\textit{(arcsec)}&\textit{(au)}&\textit{(deg.)}&&\textit{(mag)}&\textit{(M$_{\odot}$)}&\textit{(M$_{\odot}$)}&\\
\hline
\hline
\endhead
HIP 128&AC&0.98&68.9&80.6&K&3.52&1.84&0.58&0.32&20/09/2008\\
HIP 1473&$\star$&6.91&285.5&146.3&K&6.95&2.26&0.14&0.06&31/08/2009\\
HIP 2355&AB&2.01&126.3&15.9&K&4.26&2.20&0.71&0.32&16/10/2008\\
HIP 2381&--&1.77&93.9&279.1&K&6.23&2.09&0.23&0.11&26/09/2007\\
HIP 2852&--&0.93&45.4&260.6&K&5.00&1.66&0.30&0.18&17/10/2008\\
HIP 3277&--&13.06&875.4&252.5&K&2.89&2.14&0.81&0.38&29/09/2007\\
HIP 4979&$\star^{a}$&14.50&872.0&250.0&K&5.74&1.80&0.25&0.14&16/10/2008\\
HIP 4979&$\star^{a}$&15.89&955.6&254.4&K&2.72&1.80&0.81&0.45&16/10/2008\\
HIP 5300&--&0.11&5.8&126.7&K&0.57&2.10&1.64&0.78&20/09/2007\\
HIP 5310&--&0.36&16.0&175.0&K&3.72&1.83&0.39&0.21&16/10/2008\\
HIP 8847&$\star$&11.58&820.4&262.7&K&3.66&1.87&0.54&0.29&31/08/2009\\
HIP 9480&AB&0.67&24.2&297.4&K&1.20&1.93&1.17&0.61&01/09/2009\\
HIP 9480&AC&23.16&816.5&52.5&K&4.31&1.93&0.31&0.16&01/09/2009\\
HIP 11102&--&2.28&138.8&153.8&K&5.19&1.77&0.34&0.19&07/11/2005\\
HIP 11569&AaAb&0.58&23.4&43.5&K&1.91&2.19&1.15&0.53&16/10/2008\\
HIP 11569&AB&2.74&111.8&230.6&K&1.22&2.19&1.27&0.58&16/10/2008\\
HIP 11569&AC&7.25&295.2&115.3&K&2.06&2.19&1.01&0.46&16/10/2008\\
HIP 12706&AB&2.31&56.3&298.5&K&1.54&2.09&1.20&0.57&17/10/2008\\
HIP 13133&AC&7.30&473.1&71.6&K&4.27&2.59&0.20&0.08&16/10/2008\\
HIP 15353&--&4.91&269.7&17.4&K&3.91&1.71&0.40&0.23&07/11/2005\\
HIP 16292&AB&14.59&888.6&159.6&K&2.76&2.32&0.93&0.40&16/10/2008\\
HIP 17954&--&0.20&10.5&194.8&K&0.23&1.94&1.69&0.87&14/11/2008\\
HIP 18217&--&1.03&52.3&65.0&K&2.41&1.75&0.77&0.44&12/11/2008\\
HIP 18907&$\star$&16.05&575.5&182.5&K&6.12&2.35&0.28&0.12&05/02/2010\\
HIP 20542&$\star$&9.98&493.9&85.4&K&6.28&2.09&0.26&0.13&17/11/2007\\
HIP 20648&AB&1.82&83.1&341.4&K&2.56&2.11&1.03&0.49&04/02/2010\\
HIP 20713&$\star$&10.38&509.8&146.9&K&8.39&2.24&0.12&0.05&31/08/2009\\
HIP 21036&$\star$&4.76&215.2&313.6&K&4.73&1.87&0.43&0.23&05/02/2010\\
HIP 21036&$\star$&4.96&224.3&311.3&K&4.43&1.87&0.48&0.26&05/02/2010\\
HIP 23179&AB&4.87&254.8&3.5&K&1.58&2.39&1.10&0.46&17/10/2008\\
HIP 23296&AD&9.17&454.7&328.2&K&7.64&1.59&0.09&0.06&05/11/2008\\
HIP 28614&AB&0.40&19.1&22.0&K&1.00&2.19&1.68&0.77&19/12/2009\\
HIP 29711&--&4.31&280.4&239.6&K&2.62&1.80&0.73&0.41&04/11/2007\\
HIP 29852&--&0.22&13.4&210.8&K&2.00&1.95&0.94&0.48&10/11/2005\\
HIP 30419&AB&12.19&456.9&28.8&K&1.87&2.05&1.19&0.58&01/09/2009\\
HIP 31167&--&4.50&188.2&89.8&K&3.83&1.68&0.52&0.31&24/01/2002\\
HIP 33018&$\star$&2.37&137.6&295.3&K&5.43&2.50&0.70&0.28&05/02/2010\\
HIP 33018&$\star$&21.13&1224.8&297.8&K&8.46&2.50&0.19&0.08&05/02/2010\\
HIP 35350&--&9.80&303.1&33.8&K&4.09&2.39&0.61&0.26&12/04/2008\\
HIP 41375&$\star$&10.58&529.1&95.7&K&6.27&1.83&0.20&0.11&08/12/2011\\
HIP 42313&AB&2.71&133.3&262.7&K&5.56&2.59&0.44&0.17&27/01/2007\\
HIP 43584&AB&5.17&333.2&275.5&K&4.03&2.03&0.62&0.31&09/12/2011\\
HIP 44127&AB&2.39&34.7&78.4&K&4.35&1.69&0.43&0.25&05/02/2010\\
HIP 44127&AC&1.89&27.5&86.7&K&4.39&1.69&0.43&0.25&05/02/2010\\
HIP 45001&AB&17.72&1151.1&146.8&K&2.84&2.07&0.90&0.43&12/01/2005\\
HIP 45001&AC&17.51&1137.9&147.6&K&2.67&2.07&0.95&0.46&12/01/2005\\
HIP 45688&AB$^{b}$&2.62&100.4&224.5&K&1.23&2.19&1.61&0.74&12/04/2008\\
HIP 47204&--&0.70&50.4&9.1&K&0.05&1.91&1.89&0.99&18/01/2006\\
HIP 47479&--&0.10&5.7&300.2&K&0.08&2.18&2.10&0.96&15/02/2008\\
HIP 48319&--&11.78&419.8&295.4&K&5.78&2.20&0.44&0.20&12/04/2008\\
HIP 48763&--&3.63&244.8&153.0&K&2.47&1.82&0.86&0.47&08/06/2010\\
HIP 51200&--&2.44&161.3&304.4&K&3.18&1.96&0.72&0.37&04/02/2010\\
HIP 51384&--&2.08&84.3&212.4&K&4.54&1.69&0.38&0.23&12/04/2008\\
HIP 51907&$\star$&6.92&425.9&95.8&H&5.74&1.64&0.23&0.14&14/06/2008\\
HIP 51907&$\star$&7.16&440.6&95.1&H&5.66&1.64&0.25&0.15&14/06/2008\\
HIP 55266&--$^{c}$&0.90&52.9&145.0&K&5.61&2.33&0.43&0.18&23/07/2010\\
HIP 55705&--&4.98&125.6&93.1&K&2.94&1.81&0.75&0.41&05/02/2010\\
HIP 56034&AB&5.46&356.1&354.2&K&1.87&2.32&1.14&0.49&23/07/2010\\
HIP 56083&$\star$&9.17&616.4&232.4&H&6.78&1.88&0.13&0.07&14/06/2008\\
HIP 57013&$\star$&8.44&552.4&182.4&K&4.52&2.35&0.49&0.21&07/02/2005\\
HIP 57562&AD$^{c}$&2.98&176.2&293.4&K&5.90&2.24&0.25&0.11&04/01/2006\\
HIP 59923&--&8.45&464.1&281.8&H&3.34&1.94&0.68&0.35&14/06/2008\\
HIP 61498&AB&7.82&569.3&225.3&K&2.73&2.53&0.45&0.18&07/03/2005\\
HIP 64979&$\star$&10.54&663.8&167.4&K&8.17&1.86&0.10&0.05&05/02/2010\\
HIP 65241&--&0.33&20.8&197.0&K&3.06&2.05&0.64&0.31&08/02/2005\\
HIP 65477&CaCb&1.07&26.9&208.9&H&5.80&2.30&0.33&0.14&11/04/2008\\
HIP 66223&AaAb&1.38&94.7&187.7&K&5.66&1.84&0.23&0.13&13/07/2008\\
HIP 66249&--&1.79&41.0&154.1&K&6.53&2.14&0.17&0.08&05/02/2010\\
HIP 66458&AB&1.76&107.0&101.7&H&1.81&2.23&1.58&0.71&05/05/2001\\
HIP 67782&$\star$&5.28&347.0&122.0&K&5.37&1.97&0.34&0.17&05/02/2010\\
HIP 69483&--&13.55&678.7&235.5&H&2.42&2.38&1.17&0.49&11/04/2008\\
HIP 69592&--&4.06&243.2&174.7&H&5.25&1.75&0.20&0.11&12/07/2008\\
HIP 69995&$\star$&3.80&279.1&226.7&K&3.46&2.16&0.66&0.31&30/06/2004\\
HIP 70022&$\star$&1.84&116.3&53.4&H&6.45&1.84&0.18&0.10&07/06/2001\\
HIP 70400&$\star$&3.42&166.6&244.1&K&5.39&2.07&0.34&0.16&08/02/2005\\
HIP 70931&$\star$&0.60&37.6&169.6&K&4.63&1.77&0.23&0.13&30/06/2004\\
HIP 76878&AB&2.39&126.8&85.9&K&5.48&1.84&0.27&0.15&13/07/2008\\
HIP 76952&--&0.68&29.8&112.7&H&1.11&2.50&1.77&0.71&11/04/2008\\
HIP 77660&--&0.25&11.7&71.9&K&1.49&2.05&1.15&0.56&30/06/2004\\
HIP 80170&$\star$&8.33&491.8&176.1&K&6.95&2.50&0.46&0.18&05/02/2010\\
HIP 80628&AaAb&0.67&28.7&22.6&K&2.07&1.92&0.92&0.48&12/04/2008\\
HIP 80953&--&16.30&1129.8&195.6&K&2.36&2.13&1.02&0.48&24/07/2008\\
HIP 82321&AB&2.08&115.1&37.8&H&2.69&2.31&1.01&0.44&12/07/2008\\
HIP 82321&AC&1.83&101.4&33.7&H&3.07&2.31&0.91&0.39&12/07/2008\\
HIP 84012&AB&0.58&15.6&239.8&K&0.50&2.52&2.21&0.88&12/04/2008\\
HIP 84379&AB$^{c}$&12.18&280.6&285.5&K&4.56&2.15&0.46&0.21&25/07/2008\\
HIP 85822&$\star$&4.50&237.4&67.3&K&7.56&2.55&0.16&0.06&01/09/2009\\
HIP 87813&$\star$&1.88&138.8&60.2&K&6.66&2.11&0.17&0.08&27/06/2004\\
HIP 88726&--&1.75&73.4&3.6&K&0.12&1.43&1.41&0.99&02/07/2009\\
HIP 88771&AB&24.88&662.6&297.7&K&6.35&2.08&0.21&0.10&05/02/2010\\
HIP 90156&AB&3.77&212.9&348.3&H&2.05&2.23&1.10&0.49&12/07/2008\\
HIP 91919&AB&2.36&117.3&347.6&K&0.57&2.13&1.76&0.83&19/06/2008\\
HIP 91926&CD$^{d}$&2.36&112.4&259.0&K&0.00&1.99&1.99&1.00&19/06/2008\\
HIP 93506&AB&0.19&5.3&31.1&H&1.49&2.51&1.74&0.69&14/06/2008\\
HIP 93747&AB&7.27&185.1&46.6&K&4.87&2.93&0.50&0.17&17/11/2007\\
HIP 95077&$\star$&4.67&258.7&326.6&K&7.07&1.87&0.15&0.08&27/06/2008\\
HIP 95077&$\star$&4.74&262.5&321.8&K&6.98&1.87&0.16&0.09&27/06/2008\\
HIP 96313&$\star$&15.30&927.7&101.5&H&1.39&1.42&0.97&0.68&12/07/2008\\
HIP 97423&$\star$&4.64&303.7&189.7&K&4.20&1.93&0.49&0.25&18/06/2008\\
HIP 98103&--&2.83&190.4&184.4&K&5.27&2.39&0.40&0.17&18/06/2008\\
HIP 103298&AaAb&0.22&13.3&115.7&K&2.96&2.06&0.81&0.39&08/09/2008\\
HIP 104521&AB&0.80&29.2&257.4&H&2.78&1.89&0.87&0.46&14/06/2008\\
HIP 106711&--&6.89&453.6&57.0&K&8.62&2.20&0.12&0.05&08/09/2008\\
HIP 107302&$\star$&4.29&227.4&229.0&K&4.09&1.77&0.49&0.28&24/07/2008\\
HIP 109667&--&1.10&69.8&284.4&K&4.04&1.86&0.51&0.27&25/06/2010\\
HIP 109667&$\star$&5.14&326.1&181.4&K&7.49&1.86&0.10&0.05&10/09/2008\\
HIP 109857&AaAb&0.44&11.5&84.0&H&3.95&1.87&0.57&0.30&14/06/2008\\
HIP 110787&--&0.29&18.1&211.1&K&3.89&2.00&0.51&0.26&17/09/2008\\
HIP 111674&$\star$&14.74&776.0&214.8&K&5.64&2.20&0.45&0.21&08/09/2008\\
HIP 113048&AB&0.82&46.4&234.9&K&0.89&1.83&1.32&0.72&31/08/2009\\
HIP 116611&AaAb&0.95&66.6&173.1&K&5.93&2.34&0.28&0.12&29/09/2008\\
HIP 117452&AB$^{b}$&3.64&153.4&237.5&K&3.75&2.47&0.58&0.23&30/08/2009\\
HIP 118092&$\star$&0.35&23.4&328.8&K&3.05&2.04&0.71&0.35&04/01/2006\\
\hline
\multicolumn{11}{l}{$\star$ -- A newly resolved binary without a
  designation assigned within the WDS catalogue.}\\
\multicolumn{11}{l}{$a$ -- HIP 4979 B is resolved into a binary
  system itself.}\\
\multicolumn{11}{l}{$b$ -- The secondary in this pair is a
  known binary which is unresolved within the AO observations}\\
\multicolumn{11}{l}{$c$ -- The primary in this pair is a
  known binary which is unresolved within the AO observations}\\
\multicolumn{11}{l}{$d$ -- HIP 91926 CD is a wide CPM
  companion to HIP 91919 AB.}
\label{tab:ao_binaries}
\end{longtable}
\begin{longtable}{ccccccccccc}
\caption{Binary companions identified within the astrometric search}\\
HIP&CPM Companion&WDS&$\rho$&$a_{\rm proj}$&$\log a$&$\theta$&$\Delta K$&$M_1$&$M_2$&$q$\\
&&Desig.&\textit{(arcsec)}&\textit{(au)}&&\textit{(deg.)}&\textit{(mag)}&\textit{(M$_{\odot}$)}&\textit{(M$_{\odot}$)}\\
\hline
\hline
\endhead
HIP 2355&2M J00300625+2948173&$\star$ &192.1&12,057.2&4.08&355.8&8.04&2.20&0.14&0.06\\
HIP 12489&BD+26 443B&AB&28.9&2,048.9&3.31&0.8&2.87&2.32&0.93&0.40\\
HIP 19990&HD 284336&AB&180.0&5,209.7&3.72&118.4&2.45&1.41&0.47&0.33\\
HIP 21547&GJ 3305&AC&66.5&1,956.8&3.29&162.5&1.88&1.39&0.62&0.45\\
HIP 22300&TYC 3737-1375-1&$\star$ &669.1&35,033.6&4.54&142.4&2.44&1.61&0.75&0.47\\
HIP 23585&2M J05041356+4547206&AB&44.5&2,724.4&3.44&321.8&3.37&1.42&0.25&0.17\\
HIP 23875&2M J05074827-0508303&$\star$&203.5&5,575.9&3.75&11.6&5.98&2.44&0.43&0.18\\
HIP 45688&2M J09185718+3649084&$\star$&98.7&3,778.4&3.56&53.1&6.82&2.19&0.25&0.11\\
HIP 51200&2M J10273634+4135220&$\star$ &102.1&6,750.9&3.83&114.2&6.31&1.96&0.20&0.10\\
HIP 55266&HIP 55316&$\star$ &500.1&29,417.5&4.47&127.5&1.72&2.33&1.37&0.59\\
HIP 63320&2M J12585275+2809512&$\star$ &612.6&43,143.2&4.63&155.9&5.62&1.69&0.10&0.06\\
HIP 65728&HIP 65756&CA&182.6&13,033.3&4.12&110.2&1.05&2.37&1.55&0.65\\
HIP 69713&HD 234121&AB&39.0&1,133.7&3.05&32.8&1.97&1.81&0.80&0.44\\
HIP 74000&2M J15071551+1827568&$\star$ &110.3&7,967.2&3.90&321.3&2.92&2.09&0.73&0.35\\
HIP 76878&2M J15413725+1828082&AC&249.1&13,220.0&4.12&274.1&2.77&1.84&0.74&0.40\\
HIP 80883&BD+02 3118C&AC&119.0&6,316.1&3.80&169.5&4.78&2.42&0.64&0.26\\
HIP 85829&HIP 85819&AB&62.2&1,895.9&3.28&311.2&0.08&1.70&1.67&0.98\\
HIP 85922&2M J17332793-0546538&$\star$ &135.6&6,524.5&3.81&192.1&5.05&1.84&0.33&0.18\\
HIP 86263&2M J17383714-1514293&$\star$ &1,059.4&34,197.7&4.53&57.8&7.63&2.20&0.17&0.08\\
HIP 90156&HD 238865&AC&88.7&5,007.3&3.70&19.5&1.77&2.23&1.18&0.53\\
HIP 93747&2M J19065609+1340323&$\star$ &1,494.5&38,048.5&4.58&116.9&8.14&2.93&0.14&0.05\\
HIP 97421&CD-56 7835&AB&101.9&5,049.6&3.70&333.3&2.10&1.96&0.97&0.50\\
HIP 106786&2M J21372826-0755550&$\star$ &375.6&20,567.9&4.31&221.8&6.54&2.39&0.28&0.12\\
HIP 111674&TYC 3632-1527-1&$\star$ &832.6&43,822.0&4.64&190.7&4.11&2.20&0.72&0.33\\
\hline
\multicolumn{11}{l}{$\star$ -- A newly resolved binary without a
  designation assigned within the WDS catalogue.}
\label{tab:dss_binaries}
\end{longtable}
\twocolumn
\subsection{Identified companion candidates}

The companions identified both within the AO observations and from a
search for CPM companions, are plotted as a function of projected
physical separation and mass ratio in Fig. \ref{fig:completeness}. A
total of 108 companion candidates satisfying the 5 per cent
statistical criterion with mass ratios of $q\ge0.05$, corresponding
to a companion mass of 0.08 M$_{\odot}$ around a 1.5 M$_{\odot}$
primary, were identified within the high-resolution AO
observations. The observed and derived parameters 
of each resolved companion is listed in Table \ref{tab:ao_binaries}.
The companions span a range of separations between $0.08$ and
$23.2$ arcsec, and secondary masses spanning from A-type companions
($q\approx 1$) to late M-type companions at the bottom of the main
sequence ($q\approx 0.05$). Of the 113 identified companions within
the AO observations, 51 were newly resolved as a part of the VAST
survey \citep{DeRosa:2011ci,DeRosa:2012gq}, 33 of which are presented
for the first time within this paper. These 113 AO imaging companions
were complemented by an additional 24 CPM companions, the observed
and derived parameters of which are listed in
Table \ref{tab:dss_binaries}.

\subsection{Separation distribution}
\label{sec:sepdist}
\begin{figure}
\resizebox{\hsize}{!}{{\includegraphics{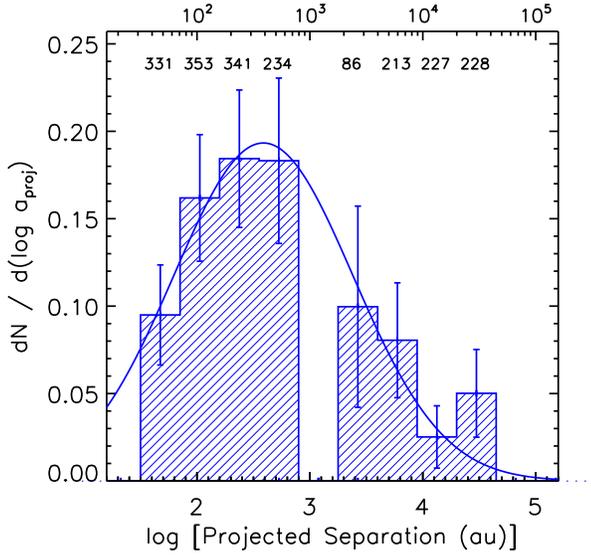}}} 
\caption{The final separation distribution was created from a
synthesis of six subsamples, minimizing the biases introduced by a
non-uniform level of completeness to companions within the adaptive
optics observations and photographic plates. The
separation distribution is constructed from two distinct data sets, 
the companion candidates resolved within the adaptive optics
observations (30 -- 800 au), and the common proper motion companions
detected within the photographic plates (1,800 -- 45,000 au). The number
of stars comprising the subsample used to estimate the frequency of
companions within each bin of the distribution are given above
the corresponding bin of the histogram. The gap in the distribution is
caused by the lack of sensitivity within the adaptive optics
observations to wide binary systems ($a_{\rm proj} \gtrsim 10^3$ au),
and the saturation of the bright targets within the digitised
photographic plates ($a_{\rm proj} \lesssim 10^3$ au). Assuming a
log-normal distribution, as measured in the separation distribution
for companions to solar-type stars \citep{Raghavan:2010gd}, the
location of the peak of the distribution was estimated   as  $\log
a_{\rm proj} = 2.59 \pm 0.13$, corresponding to $a_{\rm
proj}=387^{+132}_{-98}$ au, with a width of $\sigma_{\log a_{\rm
  proj}}=0.79\pm 0.12$.}
\label{fig:sepdist}
\end{figure}
The binary separation distribution was constructed over eight equally
wide bins in $\log a_{\rm proj}$ -- four bins spanning $\log a_{\rm
proj}=1.5-2.9$ (approx. 30 -- 800 au) for the AO companions, and four
bins spanning $\log a_{\rm proj} = 3.25-4.65$ (approx. 1,800 -- 45,000
au) for the wide CPM companions. For each bin, the subsample of
targets used to determine the frequency of companions within that bin
is the set of targets with sensitivity covering 95 per cent of the
companion phase space. Within each bin, this companion phase space is
defined by the inner and outer edge of the bin, and a mass ratio range
of $q\ge0.1$. The frequency within each bin was thus determined from
the number of companions with separations within the inner and outer
edges of the bin, resolved around targets within the subsample
described previously. As the majority of the orbital parameters for
each resolved system are not known, the separations within this study
are expressed as projected separations, without applying a correction
factor to estimate the true semi-major axis \citep{Kuiper:1935cw,Couteau:1960ts}.

The A-type star binary separation distribution is shown in Fig.
\ref{fig:sepdist}, with the number of targets used to determine the
frequency within each bin listed. Similar to previous multiplicity
surveys \citep{Duquennoy:1991wk,Raghavan:2010gd}, a
log-normal function was fitted to the measured separation distribution. The
resulting fit has a peak located at $\log a_{\rm proj} = 2.59
\pm 0.13$, corresponding to $a_{\rm proj}=387^{+132}_{-98}$ au, and
a width of $\sigma_{\log a_{\rm proj}}=0.79\pm 0.12$. The
measured separation distribution is fit well by a log-normal function,
with a slight over-abundance of CPM companions at separations of $4.25
\le \log a_{\rm proj} < 4.65$.

\subsection{Mass ratio distribution}
\begin{figure}
\resizebox{\hsize}{!}{{\includegraphics{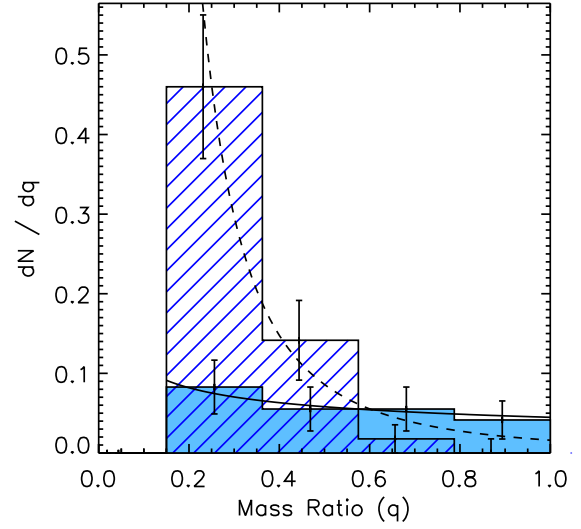}}} 
\caption{The $q$-distribution of companions resolved within
  this study with separations of 30 -- 125 au (filled
  histogram) and 125 -- 800 au (line-filled histogram). A
  power law has been fit to the two cumulative distributions, with a
power law
  index of $\Gamma = -0.5^{+1.2}_{-1.0}$ for the inner distribution
  (solid curve),  and $\Gamma = -2.3^{+1.0}_{-0.9}$ for the
  outer distribution (dashed curve). The distribution shows a
  greater abundance of lower mass companions within the outer
  distribution, with the inner distribution relatively flat for
companions with $q \ge 0.15$.}
\label{fig:2qdist}
\end{figure}

The mass ratio of each companion resolved within this study is shown
as a function of projected separation in Fig.
\ref{fig:completeness}. The apparent deficiency of $q>0.7$ companions
resolved beyond $\sim$100 au suggests that the shape of the mass
ratio ($q$-) distribution may be dependent on the projected separation
range over which it is constructed. In order to test this, two
$q$-distributions were constructed from the AO-resolved companions
between 30 -- 800 au. An inner and an outer $q$-distribution were
constructed, with the dividing separation marched from 30 to 800 au
in steps of $\log a_{\rm proj} = 0.05$. The statistical similarity of
the two distributions at each dividing separation was determined using
a two-sided  Kolmogorov-Smirnov (KS) test. A minimum value of the KS
statistic was found at a separation of $\log a_{\rm proj} = 2.1$ (125
au), with the inner (30 -- 125 au) and outer (125 -- 800 au)
$q$-distributions being statistically distinct with a KS statistic of
$3.47 \times 10^{-3}$. To minimize bias introduced by contamination
from background stars falsely identified as companions, which will be
most significant at the lowest mass ratios, only those companions with
a mass ratio of $q\ge 0.15$ were considered when constructing the
$q$-distributions.

The two resulting $q$-distributions, an inner distribution consisting
of companions with separations of $\log a_{\rm proj}=1.5-2.1$
(30 -- 125 au) and an outer distribution with $\log a_{\rm
proj}=2.1-2.9$ (125 -- 800 au), are shown in Figure \ref{fig:2qdist}.
The inner $q$-distribution, comprising 18 companions resolved
around 341 targets, is consistent with a flat distribution, while the
outer $q$-distribution, comprising 35 companions resolved around 266
targets, shows a significant increase in companion fraction as a
function of decreasing mass ratio.

\subsection{Companion star and multiplicity fractions}
\begin{table*}
\centering
\caption{Companion star and multiplicity fractions from various surveys}
\begin{tabular}{ccccc}
Spectral&Type&Separation range&Companion mass&Value\\
Type&&\textit{$\log$ [au]}&range&\textit{per cent}\\
\hline
\hline
A&CSF&$1.5 \le \log a_{\rm proj} < 2.9$&$q \ge 0.10$&$21.9\pm2.6$\\
A&CSF&$1.5 \le \log a_{\rm proj} < 2.9$&$q \ge 0.05$&$\ge 26.0$\\
A&CSF&$1.5 \le \log a_{\rm proj} < 4.0$&$q \ge 0.10$&$33.8\pm2.6$\\
A&CSF&\textit{all}&M$_2 \ge 0.08$M$_{\odot}$&$68.9\pm7.0$\\
A&MF&\textit{all}&M$_2 \ge 0.08$ M$_{\odot}$&$\ge 43.6 \pm 5.3$\\
\hline
BA (ScoCen)&CSF&$1.5 \le \log a_{\rm proj} < 2.9$&$q \ge
0.10$&$25.1\pm3.6$\\
BA (ScoCen)&CSF&$1.5 \le \log a_{\rm proj} < 2.9$&$q \ge 0.05$&$\ge
28.1$\\
\hline
FGK&CSF&$1.5 \le \log a_{\rm proj} < 2.9$&$q \ge 0.10$&$19.6\pm2.1$\\
FGK&CSF&$1.5 \le \log a_{\rm proj} < 4.0$&$q \ge 0.10$&$27.8\pm2.5$\\
FGK&CSF&\textit{all}&M$_2 \ge 0.08$ M$_{\odot}$&$61.0\pm3.7$\\
FGK&MF&\textit{all}&M$_2 \ge 0.08$ M$_{\odot}$&$46.0 \pm 2.0$\\
\hline
M&CSF&$1.5 \le \log a_{\rm proj} < 2.9$&M$_2 \ge 0.08$ M$_{\odot}$&$17.1\pm5.4$\\
M&CSF&$1.5 \le \log a_{\rm proj} < 4.0$&M$_2 \ge 0.08$ M$_{\odot}$&$24.6\pm6.5$\\
M&MF&\textit{all}&M$_2 \ge 0.08$ M$_{\odot}$&$42.0 \pm 9.0$\\
\hline
LT&CSF&$\log a_{\rm proj} \ge 1.6$&M$_2 \ge 0.03$ M$_{\odot}$&$\le
2.3$\\
LT&MF&\textit{all}&$q \ge 0.20$&$12.5 \pm 3.0$\\
\hline
\end{tabular}
\label{tab:fraction}
\end{table*}
\begin{table*}
\centering
\caption{Companions listed within the literature used in the
  calculation of the lower limit of the multiplicity fraction}
\begin{tabular}{ccccc|ccccc}
HIP&WDS&$\log P_d$&$\rho$&Reference&HIP&WDS&$\log P_d$&$\rho$&Reference\\
&Desig.&&(arcsec)&&&Desig.&&(arcsec)\\
\hline
\hline
128&AB&0.98&--&\citet{Carquillat:2003di}&54746&--&--&3.80&\citet{Tokovinin:2012kb}\\
2578&AaAb&--&0.10&\citet{vandenBos:1927uz}&55266&--&0.41&--&\citet{Lloyd:1981vs}\\
3414&--&0.29&--&\citet{Mannino:1955ve}&57562&AaAb&--&0.40&\citet{McAlister:1989hf}\\
6514&--&1.55&--&\citet{Fekel:2011cl}&57606&--&4.69&--&\citet{Brendley:2006vw}\\
6686&--&2.88&--&\citet{Samus:2009tf}&57646&--&0.44&--&\citet{Petrie:1926wa}\\
8588&AB&--&178.00&\citet{Tokovinin:2012jg}&58001&AB&--&20276.12&\citet{Shaya:2010dt}\\
8903&--&2.03&--&\citet{Pourbaix:2000ef}&58758&--&1.39&--&\citet{Moore:1931tl}\\
9153&AB&--&37.10&\citet{Halbwachs:1986uo}&61932&AB&4.49&0.50&\citet{Malkov:2012cq}\\
9480&AC&--&23.20&\citet{Mason:2001ff}&61932&AD&--&2719.24&\citet{Shaya:2010dt}\\
9836&--&1.18&--&\citet{Jones:1931uk}&61937&AC&--&5590.10&\citet{Shaya:2010dt}\\
10064&--&1.50&--&\citet{Pourbaix:2000ef}&62983&--&--&$<$ 0.10&\citet{Africano:1975ig}\\
11486&--&--&0.10&\citet{Horch:2008et}&64692&--&--&0.60&\citet{McAlister:1993ez}\\
11569&CaCb&--&0.40&\citet{Christou:2006ip}&69483&BaBb&3.25&--&\citet{Kiyaeva:2006ta}\\
12706&AC&--&843.10&\citet{Alden:1924fa}&69483&AC&--&108.80&\citet{Tokovinin:2012jg}\\
12828&--&3.08&--&\citet{Abt:1965fz}&69974&--&2.32&--&\citet{Stickland:1990ua}\\
13133&AaAb&0.08&--&\citet{Duerbeck:1979uz}&70931&Aa1Aa2&1.07&--&\citet{Kaufmann:1973va}\\
15197&--&1.25&--&\citet{Abt:1985bs}&71075&AaAb&--&0.10&\citet{Morgan:1978tn}\\
15197&--&--&1383.23&\citet{Shaya:2010dt}&72622&AD&--&8977.46&\citet{Shaya:2010dt}\\
16591&--&-0.04&--&\citet{Rucinski:2005gu}&74000&--&3.47&0.10&\citet{Eggen:1946ia}\\
19893&--&--&--&\citet{Samus:2009tf}&75695&--&3.58&--&\citet{Neubauer:1944is}\\
20087&--&3.62&--&\citet{Pourbaix:2000ef}&76852&AB&3.90&0.10&\citet{Muterspaugh:2010cl}\\
20713&--&3.72&--&\citet{Abt:1965fz}&76996&AB&--&65.90&\citet{Mason:2001ff}\\
20894&--&2.15&--&\citet{Torres:1997gv}&77233&AD&--&1643.04&\citet{Shaya:2010dt}\\
21029&--&--&249.80&\citet{Peterson:1981er}&80628&--&1.43&--&\citet{Gutmann:1965tp}\\
21039&--&1.77&--&\citet{Debernardi:2000wq}&80628&AB&--&18227.26&\citet{Shaya:2010dt}\\
21273&&2.69&--&\citet{Abt:1965fz}&80883&AB&4.67&--&\citet{Heintz:1993ft}\\
21402&Aa1Aa2&0.55&--&\citet{Lane:2007jh}&84379&AaAb&--&0.10&\citet{Bonneau:1980tw}\\
21402&Ab1Ab2&0.90&--&\citet{Lane:2007jh}&84606&--&--&0.92&\citet{Horch:2011de}\\
21402&AaAb&3.82&--&\citet{Lane:2007jh}&85829&AaAb&1.58&--&\citet{Margoni:1992wb}\\
21547&CaCb&--&0.20&\citet{Kasper:2007dm}&86263&AaAb&--&0.30&\citet{Isobe:1990vs}\\
21644&--&--&0.10&\citet{Horch:2004to}&86263&Aa1Aa2&0.36&--&\citet{Young:1910ei}\\
21673&--&1.59&--&\citet{Abt:1985bs}&87212&AaAb&--&0.10&\citet{McAlister:1987bk}\\
22287&AaAb&--&0.40&\citet{Horch:2011de}&87212&AB&--&209.10&\citet{Lepine:2007cf}\\
23296&AaAb&0.91&--&\citet{Fekel:2006ca}&89925&--&0.74&--&\citet{Fekel:2009cc}\\
23983&--&2.19&--&\citet{Debernardi:2000wq}&90156&AaAb&--&--&\citet{Frost:1924jc}\\
24340&--&--&0.10&\citet{Mason:1999ku}&90156&CaCb&0.43&--&\citet{Halbwachs:2012es}\\
26309&AB&--&1211.97&\citet{Shaya:2010dt}&91919&ABCD&--&208.80&\citet{Shaya:2010dt}\\
26563&--&2.65&--&\citet{Abt:1965fz}&91971&AaAb&0.63&--&\citet{Abt:1985bs}\\
28614&AaAb&0.65&--&\citet{Muterspaugh:2008fg}&91971&AD&--&41.10&\citet{Shaya:2010dt}\\
28614&BaBb&0.68&--&\citet{Muterspaugh:2008fg}&92024&BaBb&--&0.20&\citet{Biller:2007ht}\\
29850&AaAb&3.53&--&\citet{Hartkopf:1996cw}&92024&AB&--&71.40&\citet{Mason:2001ff}\\
30060&--&--&--&\citet{Samus:2009tf}&98103&AB&0.52&--&\citet{Lucy:1971ky}\\
33202&AB&5.85&7.40&\citet{Hopmann:1974ta}&101093&--&2.92&--&\citet{Abt:1962fo}\\
33202&AD&--&152.50&\citet{Tokovinin:2012jg}&101800&--&1.04&--&\citet{Harper:1935ty}\\
41081&AB&--&1119.82&\citet{Shaya:2010dt}&106786&--&3.90&--&\citet{Abt:1965fz}\\
42806&AaAb&--&--&\citet{Lee:1910hf}&107556&--&0.01&--&\citet{Batten:1992vt}\\
43970&AB&--&975.90&\citet{Shaya:2010dt}&113048&AaAb&1.38&--&\citet{Margoni:1992wb}\\
45688&BaBb&--&0.20&\citet{McAlister:1993ez}&113996&AB&3.90&--&\citet{Hartkopf:1996cw}\\
54214&--&--&0.60&\citet{Tokovinin:2012kb}&116611&Aa1Aa2&-0.30&--&\citet{Rucinski:2005gu}\\
\hline
\end{tabular}
\label{tab:literature_binaries}
\end{table*}

There are two different quantities which can be used to express the
fraction of stars within multiple systems; the multiplicity fraction
(MF; \citealp{Reipurth:1993ux}), defined as
\begin{equation}
{\rm MF}=\frac{B+T+Q \dots}{S+B+T+Q \dots}
\end{equation}
and the CSF \citep{Goodwin:2004cn}, defined as
\begin{equation}
{\rm CSF}=\frac{B+2T+3Q \dots}{S+B+T+Q \dots}
\end{equation}
where $B$, $T$ and $Q$ are the number of binary, triple and
quadruple systems, respectively. From our AO observations,
the CSF between 30 -- 800 au (CSF$_{\rm 30 - 800 au}$) was
calculated by summing the inner four bins of the separation
distribution in Fig. \ref{fig:sepdist}, leading to a CSF of $21.9
\pm 2.6$ per cent for companions with $q\ge 0.1$. As a large-subset of
the AO observations were sensitive to stellar companions with $q<0.1$,
a lower limit of the CSF$_{\rm 30 - 800 au}$ value including all
stellar companions with $q \ge 0.05$ was estimated to be 26.0 per
cent. The CSF over a wider separation range (30 -- 10,000 au; CSF$_{\rm 30 - 10,000 au}$),
calculated by integrating the fit to the separation distribution shown
in Fig. \ref{fig:sepdist}, was estimated to be $33.8 \pm 2.6$ per
cent. These three estimates of the CSF are given in Table
\ref{tab:fraction}, alongside the separation range over which they
were constructed, and the mass range of companions
included. 

The total CSF for A-type stars, considering companions at all possible
separations, can be estimated by combining the CSF$_{\rm 30 - 10,000
  au}$ measurement from this study with the results of previous
spectroscopic surveys. These surveys typically measured the frequency
of companions to three categories of A-type stars; normal
($30.9\pm7.5$ per cent; \citealp{Abt:1965fz}), metallic-lined
($63.7\pm8.4$ per cent; \citealp{Carquillat:2007hj}) and chemically
peculiar ($43.0\pm6.0$ per cent; \citealp{Carrier:2002hs}). When
weighted according to the fraction of the VAST sample within each of
these categories (normal - 85.3 per cent, Am - 11.5 per cent, Ap - 3.2
per cent), the weighted frequency becomes $35.1 \pm 6.5$ per cent. The
total CSF for A-type stars was then calculated as the sum of the
CSF$_{\rm 30 - 10,000 au}$ measurement from this study, and the
weighted frequency from the spectroscopic surveys, a total of $68.9 \pm 7.0$ per
cent (Table \ref{tab:fraction}). Although not significantly higher
than the value for solar-type primaries, measured to be $61.0 \pm 3.7$ per cent \citep{Raghavan:2010gd}, the
completeness of the spectroscopic surveys has not been assessed, and
binaries with separations of the order of 10 au may have been missed
due to their small radial velocity variations over the time period in
which they were observed spectroscopically.

Additionally, a lower limit of the MF over all companion
separations was estimated for the sample of 156 A-type stars which
both have AO observations and were searched for wide CPM
companions, providing sensitivity to companions over the widest
separation range, and crucial for the best assessment of the frequency of higher
order multiples (described in Section 7.6). This lower limit of $43.6 \pm 5.3$ per cent, listed in Table
\ref{tab:fraction}, was calculated using the companions reported
within Tables \ref{tab:ao_binaries} and \ref{tab:dss_binaries},
combined with physically associated binaries recorded within the
Washington Double Star Catalog (WDS; \citealp{Mason:2001ff}), the Ninth Catalogue of
Spectroscopic Binary Orbits (SB9; \citealp{Pourbaix:2004dg}) and eclipsing
binaries within the General Catalogue of Variable Stars (GCVS;
\citealp{Samus:2009tf}), which are all listed in Table
\ref{tab:literature_binaries}.

\section{Discussion}
\subsection{Comparison samples of different masses and ages}
In order to place the results of this volume-limited
multiplicity of A-type stars into context, it is necessary to consider
samples against which the results will be compared. The results of the
VAST survey are compared to volume-limited multiplicity surveys of
lower mass solar-type, M-dwarf and brown dwarf primaries
\citep{Fischer:1992cj, Reid:2008gr, Raghavan:2010gd}, allowing for an
investigation of the various multiplicity statistics as a function of
primary mass. The results cannot be easily compared to more massive
stars within the field given the very small number of stars with
spectral type earlier than B7 within 75 pc \citep{Abt:2011ha}. The
VAST results are also compared with B- and A-type stars within the
nearby ScoCen OB association \citep{Kouwenhoven:2005jf}, allowing for
a comparison of the multiplicity statistics between cluster and field
populations. For each of these comparison samples, the separation and
mass ratio of each companion is known, so a fair comparison can be
made between the various surveys over a common range of companion
separation and mass ratios (Table \ref{tab:fraction}). More recent
studies of M-dwarf multiplicity have not been included as comparison
samples as these surveys do not have sensitivity to companions with
separations $\gtrsim 200$ au \citep{Bergfors:2010hm, Janson:2012dc}.

\subsection{Multiplicity as a function of primary mass}
\subsubsection{Observed trend}
\begin{figure}
\resizebox{\hsize}{!}{{\includegraphics{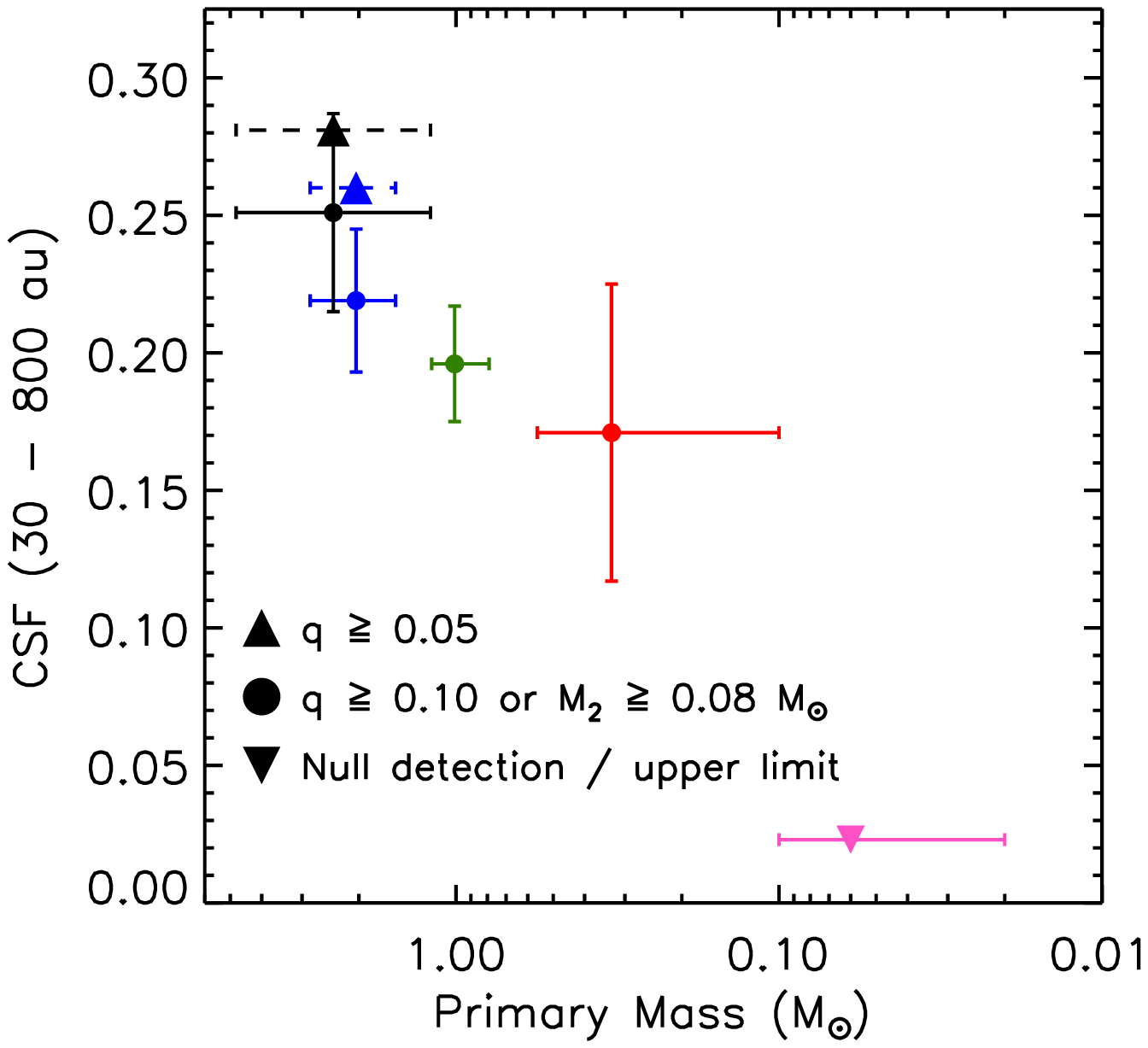}}} 
\caption{The CSF$_{\rm 30 - 800 au}$ measured within the separation
range 30 -- 800 au measured for, from left to right, Sco OB2 primaries
  (black point, \citealp{Kouwenhoven:2005jf}), field A-type stars (blue point, this study), field
  solar-type stars (green point, \citealp{Raghavan:2010gd}), field M-dwarf
  primaries (red point, \citealp{Fischer:1992cj}). The upper limit for companions to
  brown dwarfs within this separation range is also shown
  (pink downward triangle, \citealp{Allen:2007hd}). A lower limit on
the CSF$_{\rm 30 - 800 au}$ including all stellar
  companions to A-type stars ($q\gtrsim 0.05$) was estimated by
  including companions resolved within this study with mass ratios
  within the range $0.05 \le q < 0.1$ (blue upward pointing
  triangle, dashed error bars). Similarly, for the study of Sco OB2
  primaries, the CSF  including all stellar companions is plotted (black
  upward pointing triangle, dashed error bars).}
\label{fig:multiplicity}
\end{figure}
\begin{figure}
\resizebox{\hsize}{!}{{\includegraphics{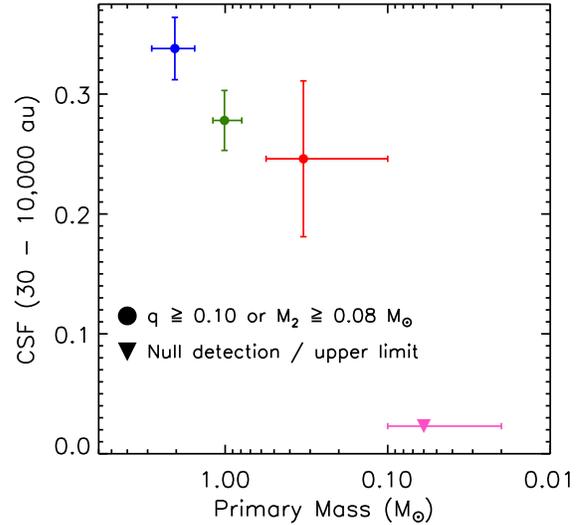}}} 
\caption{The CSF$_{\rm 30 - 10,000 au}$ is estimated from the
  log-normal fit to the separation distribution within the range 30
-- 10,000 au for, from left to right, field A-type stars (this study,
blue point), solar-type stars (green point,
\citealp{Raghavan:2010gd}) and field M-dwarf primaries (red point,
\citealp{Fischer:1992cj}). The upper limit for companions to
  brown dwarfs within this separation range is also shown
  (pink downward triangle, \citealp{Allen:2007hd}).}
\label{fig:multiplicity_fit}
\end{figure}
The A-type star CSF was calculated and compared with other samples over
two separation ranges -- the $30 - 800$ au range continuously covered
by the AO data, and the $30 - 10,000$ au range which was estimated from
the fit to the separation distribution shown in Fig. \ref{fig:sepdist}. The
CSF$_{\rm 30 - 800 au}$ measured within the VAST survey is plotted
alongside the observed CSF$_{\rm 30 - 800 au}$ for solar-type
\citep{Raghavan:2010gd}, M-dwarf \citep{Fischer:1992cj} and B- and
A-type Sco OB2 \citep{Kouwenhoven:2005jf} primaries in Fig.
\ref{fig:multiplicity}. The CSF$_{\rm 30 - 800 au}$ values for these
three literature surveys were restricted to only include companions
with a mass ratio of $q\ge 0.1$, and projected separations between 30
and 800 au, ensuring a fair comparison between all surveys with a
uniform mass ratio limit. For the surveys of nearby solar-type stars
and Sco OB2 primaries, the CSF$_{\rm 30 - 800 au}$ value could be
calculated directly from the table of companions presented within
each study. For the survey of nearby M-dwarfs, the CSF$_{\rm 30 - 800
au}$ value was estimated from an integration of the frequency of
companions per primary per au given over several discrete separation
ranges \citep{Fischer:1992cj}. In addition to the stellar results, the upper limit to the
L-dwarf CSF beyond 40 au is plotted \citep{Allen:2007hd}. The measured
CSF for A-type primaries is similar to that of early-type stars within
the young Sco OB2 stellar association \citep{Kouwenhoven:2005jf}, when
limited to a common separation and mass ratio range (Fig.
\ref{fig:multiplicity}).

While the uniform $q\ge 0.1$ limit reaches the bottom of the main
sequence for the solar-type and lower mass samples, this limit does
not include all stellar companions to A-type stars. To compare the
CSF$_{\rm 30 - 800 au}$ values with a common companion mass limit,
the $q$ limit for the A-type stars needs to be extended to include
systems with $q\ge 0.05$. For the VAST and Sco Cen OB2 samples, the
CSF$_{\rm 30 - 800 au}$ was calculated with a limit of $q\ge
0.05$, and the values are listed in Table \ref{tab:fraction} and
plotted in Fig. \ref{fig:multiplicity}. These values represent
lower limits to the CSF$_{\rm 30 - 800 au}$, since the data are not
uniformly sensitive to companions at these extreme mass ratios.

Extending the range of companion separations considered, the CSF$_{\rm
30 - 10,000 au}$ values were estimated from the fit to the observed
companions to solar-type \citep{Raghavan:2010gd} and M-dwarf
\citep{Fischer:1992cj} primaries, and are listed in Table
\ref{tab:fraction} and plotted in Fig. \ref{fig:multiplicity_fit}. A
similar value was not calculated for the Sco OB2 primaries, as the
observations were only sensitive to companions with separations within
1600 au. The increase in the CSF as a function of increasing primary
mass is apparent over both separation ranges, albeit within the large
uncertainty of the CSF for M-dwarf primaries (Figs
\ref{fig:multiplicity} and \ref{fig:multiplicity_fit}).

\subsubsection{Theoretical predictions}
\begin{figure}
\resizebox{\hsize}{!}{{\includegraphics{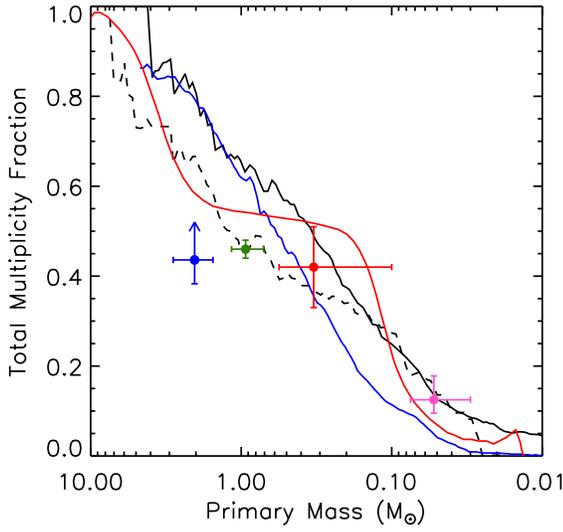}}} 
\caption{The measured multiplicity
  fraction of four stellar clusters simulations as a function of
  primary mass. The results of barotropic and radiation hydrodynamical
  simulations are plotted as black solid and dashed lines,
  respectively \citep{Bate:2009br,Bate:2011hy}, a numerical
  Monte Carlo calculation looking at small-$N$ interactions are plotted
  as a solid blue line (clusters with $N<10$ stars; \citealp{Sterzik:2003jh}), and an ensemble of
  small $N$-body clusters are plotted as a solid red line
  (a ring of $N=6$ stars, with mass dispersion of $\sigma_{\log
    M}=0.2$; \citealp{Hubber:2005it}). Over-plotted for reference are,
  from left to right, the lower limit on the multiplicity fraction of
  A-type stars measured within this survey, and the observed
  multiplicity fraction of nearby solar-type \citep{Raghavan:2010gd},
  M-dwarf \citep{Fischer:1992cj} and brown dwarf \citep{Reid:2008gr}
  primaries.}
\label{fig:multiplicity_simulations}
\end{figure}
Hydrodynamical \citep{Bate:2009br,Bate:2011hy} and numerical $N$-body
\citep{Durisen:2001kq,Sterzik:2003jh,Hubber:2005it} simulations of stellar clusters
predict that the total multiplicity increases with increasing primary
mass with different functional forms, as shown in Fig.
\ref{fig:multiplicity_simulations}. Both types of models reproduce
the observed trend of multiplicity, however some discrepancies
are present when the predictions are examined over a specific mass
range. For example, the tight constraint on the measurement of the MF
of solar-type primaries is lower than the predictions emerging
from the barotropic hydrodynamical simulation \citep{Bate:2009br},
and numerical $N$-body interactions \citep{Sterzik:2003jh,Hubber:2005it}, while
being consistent with the radiation hydrodynamical simulation
\citep{Bate:2011hy}.  As there was not uniform sensitivity to stellar
companions to A-type primaries over the full range of companion separations,
only a lower limit to the MF was estimated. Combining the results of
the VAST survey with known binaries resolved in previous
interferometric, spectroscopic and AO imaging surveys,
leads to a lower limit on the total MF shown in Fig.
\ref{fig:multiplicity_simulations}. The comparison to the various
simulations would suggest a significant number of companions interior
to $\sim$30 au remain unresolved.

\subsection{Separation distribution}
\subsubsection{Comparison with previous observations}
\begin{figure}
\resizebox{\hsize}{!}{{\includegraphics{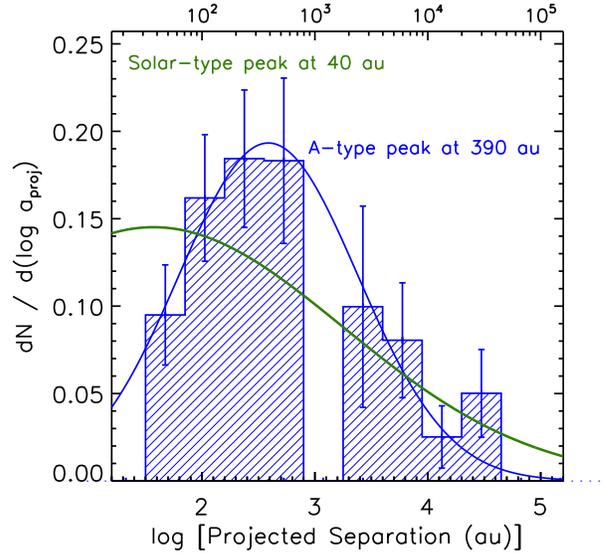}}} 
\caption{The separation distribution (blue histogram) and
  corresponding log-normal fit (blue curve), are plotted alongside
  the log-normal fit to the separation distribution of companions resolved within the
  \citet{Raghavan:2010gd} survey (green curve).}
\label{fig:sepdist_withG}
\end{figure}
The trend of a wider separation peak to the distribution as a
function of increasing primary mass is consistent with previous
multiplicity surveys of brown dwarf \citep{Burgasser:2006hd} and M-dwarf
\citep{Fischer:1992cj} primaries (Figs \ref{fig:sepdist_withG} and
\ref{fig:simulation_peak}). The log-normal fit to the measured
distribution is also significantly narrower than the solar-type
companion separation distribution, with a standard deviation of
$\sigma_{\log a}=0.79\pm 0.12$ compared to $\sigma_{\log a}\approx
1.68$ for solar-type companions (Fig. \ref{fig:sepdist_withG}), although the fit to the
solar-type separation distribution included companions at all
separations. Extrapolating the fit to the A-type star companion separation
distribution to closer separations suggests a complete lack of companions
with separations of $\log a_{\rm proj} < 0$, inconsistent with known
spectroscopic companions to normal and metallic-lined A-type stars
\citep{Abt:1965fz,Carquillat:2003di}. These spectroscopic systems are not
included within the separation distribution plotted in Fig.
\ref{fig:sepdist}, as the companion mass ratio sensitivity limits are
not quantified. It is evident that further investigations are required
in order to fully constrain the shape of the separation distribution,
with wide-field direct imaging filling the gap in the separation
distribution between the AO observations and photographic plates
presented within this work, and observations with high-order AO
systems providing sensitivity to companions of mass ratio $q \ge 0.1$
with projected separations between 10 and 100 au. For companions at
separations interior to the angular resolution of AO instruments,
radial velocity monitoring provides the only method sensitive enough
to detect companion with mass ratios as extreme as $q=0.1$, with
current-generation interferometric techniques restricted by their
limited dynamical range.

\subsubsection{Comparison with theoretical models}
\begin{figure}
\resizebox{\hsize}{!}{{\includegraphics{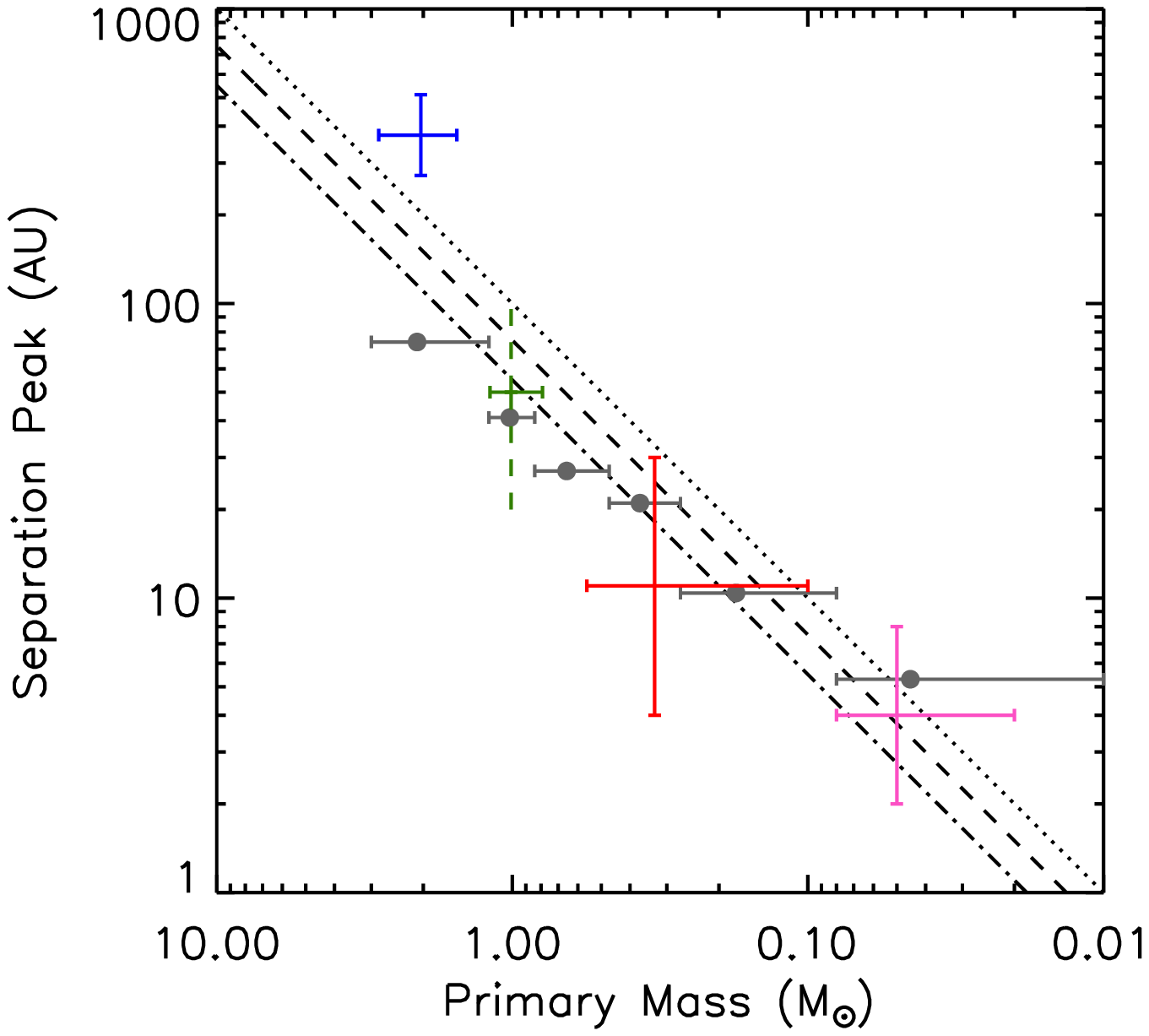}}} 
\caption{The location of the peak of the separation
  distribution as a function of primary mass for, black data points
  from left to right; A-type (this study), solar-type
  \citep{Raghavan:2010gd}, M-dwarf \citep{Fischer:1992cj} and brown
  dwarf \citep{Burgasser:2006hd} primaries. As the
  uncertainty in the location of the peak of the separation
  distribution was not provided by \citet{Raghavan:2010gd}, the
  corresponding vertical error bar is dashed. The observations show a
  clear trend of an increase in the peak of the distribution as a
  function of primary mass, similar to the theoretical predictions 
  from dynamical simulations (grey points; \citealp{Sterzik:2003jh}),
  and from models of secondary fragmentation (black lines;
  \citealp{Whitworth:2006cd}).}
\label{fig:simulation_peak}
\end{figure}

Simulations of dynamical interactions within stellar clusters and
numerical calculations of companion formation through disc
fragmentation both predict an increase in the location of the peak of
the separation distribution as a function of increasing primary mass,
and the different predictions are shown in Fig.
\ref{fig:simulation_peak} \citep{Sterzik:2003jh, Whitworth:2006cd}.
These results are consistent with hydrodynamical simulations showing
stellar binaries have a wider median separation ($\sim$26 au) than
brown dwarf binaries ($\sim$10 au), although the latter result may be
affected by the resolution limit of the calculation
\citep{Bate:2009br}.

Measurements of the position of the peak of the companion separation
distribution for lower mass primaries are consistent with these
predictions \citep{Fischer:1992cj, Burgasser:2006hd, Raghavan:2010gd},
while the A-type star separation peak occurs at a wider separation than
expected (Figure \ref{fig:simulation_peak}). Although this peak occurs
at a significantly wider separation than that of resolved
planetary-mass companions to nearby A-type stars
\citep{Kalas:2008cs, Marois:2008ei, Lagrange:2009hq}, the dynamical
interaction between a disc and a companion at the typical separation
of $\sim$390 au may lead to a truncation of the disc to a radius as small
as $\sim$50 au \citep{Artymowicz:1994eu}. Interior to the disc
truncation radius, the perturbations induced by the companion may
significantly affect the planet formation process
\citep{Nelson:2000id,Kley:2008dc}. This would suggest that a
majority of A-type star binaries, and as such a significant minority
of A-type stars in general, may not be amenable to the formation of
planetary-mass companions.

\subsection{Mass ratio distribution}
\subsubsection{Comparison with previous observations}
\begin{figure}
\resizebox{\hsize}{!}{{\includegraphics{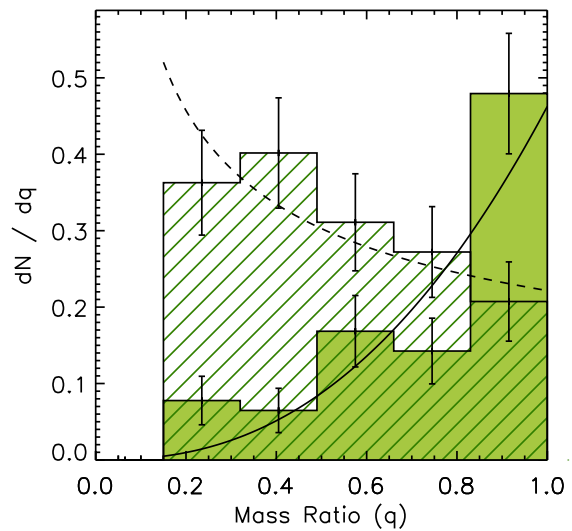}}} 
\caption{The $q$-distribution of companions resolved to
solar-type stars, with separations $\le 30$ au (filled histogram), and
those with separations $> 30$ au (line-filled histogram). The
separation and mass ratio of each component were obtained from fig. 
11 of \citet{Raghavan:2010gd}. The two
populations are statistically distinct, with a KS statistic of
$10^{-8}$. A power law has been fit to the two
cumulative distributions, with a power law index of
$\Gamma=2.5^{+0.9}_{-1.3}$ for the inner distribution (solid curve),
and $\Gamma=-0.5^{+0.5}_{-0.4}$ for the outer distribution (dashed
curve).}
\label{fig:gqdist}
\end{figure}

\begin{figure}
\resizebox{\hsize}{!}{{\includegraphics{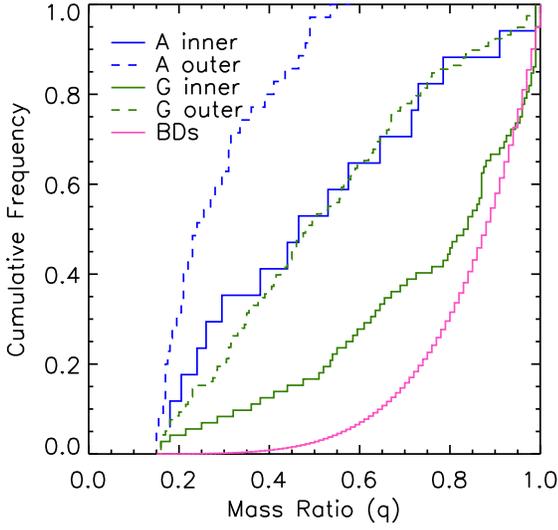}}} 
\caption{The cumulative $q$-distribution for companions
  resolved around A-type primaries, with separations of 30 -- 125
au (blue solid histogram -- inner distribution) and 125 -- 800 (blue
dashed histogram -- outer distribution). The
  greater frequency of lower mass companions within the outer
  distribution is apparent. For solar-type primaries, a similar
  division into an inner and outer distribution was performed, with
  those companions with separations of $\le 30$ au (green solid
histogram -- inner distribution) and $>30$ au (green dashed histogram --
outer distribution).  Each cumulative
  $q$-distribution only include those companions with mass ratio of $q
  \ge 0.15$. The functional form of the observed $q$-distribution for
companions to brown dwarf primaries is also
  plotted, considering all companions resolved with separations $\ge
2$ au (pink solid histogram, \citealp{Burgasser:2006hd}).}
\label{fig:qdist_withGK}
\end{figure}
\begin{figure}
\resizebox{\hsize}{!}{{\includegraphics{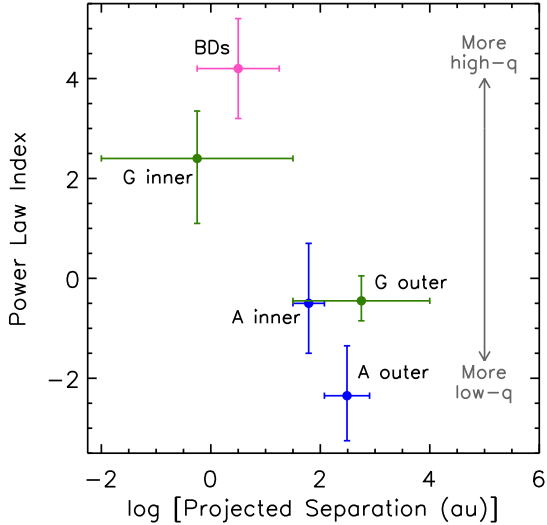}}} 
\caption{The power law index best fit to the cumulative
$q$-distributions presented in Figure \ref{fig:qdist_withGK}, for
A-type
  primaries (blue points), solar-type primaries (green points) and
  brown dwarf primaries (pink point). The data are indicative of a
  trend of a greater frequency of equal-mass companions at closer
  separations, for both A-type and solar-type primaries.}
\label{fig:sepvgamma}
\end{figure}
Determining how the $q$-distribution changes as a function of
separation for different primary masses will allow for a greater
insight into the various formation processes for binary companions
which may dominate over different separation regimes. As the
$q$-distribution for companions to A-type stars has been divided into
two statistically distinct subsamples (Fig. \ref{fig:2qdist}); a
similar technique was applied to the companions resolved around
solar-type stars \citep{Raghavan:2010gd}. The inner and outer
subsamples were found to be most statistically distinct at a dividing
separation of $\log a = 1.5$ ($\sim$30 au), with a KS statistic of
$9.92 \times 10^{-9}$. The inner and outer $q$-distributions are plotted in
Fig. \ref{fig:gqdist}, showing that lower mass companions to
solar-type primaries are preferentially found in wider orbits. The
cumulative $q$-distribution of these two distinct subsamples
are plotted in Fig. \ref{fig:qdist_withGK}, alongside the two
cumulative distributions measured for companions to A-type primaries, and the
functional form of the cumulative $q$-distribution of
companions to brown dwarf primaries \citep{Burgasser:2006hd}.
By fitting a power law to each of the five distributions, the
best-fitting power-law index was estimated and plotted as a function of
separation (Fig. \ref{fig:sepvgamma}). This shows a trend of a
greater frequency of lower mass companions at wider separations for
both A-type and solar-type primaries, although the A-type comparison
is made over a significantly narrower separation range. The
cumulative $q$-distribution for companions to M-dwarf primaries
is not plotted, although the results of recent surveys are consistent
flat distribution between 1 and 200 au
\citep{Bergfors:2010hm,Janson:2012dc}.

\subsubsection{Comparison with theoretical models}
Two binary formation scenarios are thought to pre-dominate over the 30
-- 10,000 au separation range to which this study is complete, the
initial fragmentation of a pre-stellar molecular cloud
(e.g. \citealp{Boss:1979jk,Bonnell:1991gt}) and fragmentation of
a large circumstellar disc (e.g.
\citealp{Adams:1989ca,Bonnell:1994vx,Woodward:1994eg}). The initial
fragmentation of a cloud prior to the formation of protostellar
objects can produce binary systems with separations ranging between
$10^1$ and $10^4$ au, with a wide range of companion mass ratios
\citep{Boss:1986kj,Bonnell:1991gt,Bonnell:1992ep,Bate:1995wm}. A
scale-free fragmentation model, in which the companion
$q$-distribution is independent of the initial clump mass
\citep{Clarke:1996tl}, can be tested against the observations
presented here. The initial fragmentation model predicts an MF which is weakly dependent on primary 
mass, and a mass ratio distribution which is independent, or weakly
dependent on primary mass. These predictions are inconsistent with
both the observed trend in multiplicity as a function of primary
mass for stellar primaries (Figs \ref{fig:multiplicity} and
\ref{fig:multiplicity_fit}), and the observed variation in the
companion $q$-distribution between companions to solar-type and
A-type primaries (Fig. \ref{fig:qdist_withGK}). Star formation
within a more clustered environment may introduce a dependence on
primary mass, with more massive primaries having $q$-distributions
skewed towards less massive
companions \citep{Bonnell:1992ep,Bate:2001td}. Simulations of the
dynamical decay of small clusters, formed through the fragmentation of
an initial cloud \citep{Sterzik:1998vo}, have produced companion
$q$-distributions which are consistent with observed companion
$q$-distributions, and its dependence on primary mass.

Subsequent to the fragmentation of an initial cloud the conservation
of angular momentum causes the infall of material from the
surrounding cloud to form a protostellar disc \citep{Bonnell:1994vx}.
These discs can fragment to produce substellar and stellar
companions, provided that a mechanism for the disc to become
gravitationally unstable, and subsequently cool efficiently, is
present (e.g. \citealp{Kratter:2009gs}). This formation process is
thought to be more important for more massive stars
\citep{Kratter:2011ul}, due primarily to the large reservoir of
material within the massive discs of these stars (e.g.
\citealp{Fukagawa:2010vp}). This process may lead to a significant
number of disc-born lower mass companions to more massive stars
\citep{Kratter:2010gf,Stamatellos:2011ev}, with the preferential
equalisation of the mass ratios of those binaries
formed at close separations \citep{Bate:1997uf,Bate:2000ir}. 

The measured $q$-distribution presented within this study, with a
greater abundance of lower mass companions around more massive stars,
is consistent with predictions from both initial and disc
fragmentation. The shape of the $q$-distribution for companions
resolved between 30 to 800 au is consistent with a population of
disc-born companions \citep{Kratter:2010gf}, and such
companions are within a separation range coincident with the extent of
known circumstellar discs of pre-main sequence A-type stars
(e.g. \citealp{Dent:2006gr,Hamidouche:2006jh,Fukagawa:2010vp}). For
the widest companions, formation via disc fragmentation is not
possible, and the observed frequency of wide ($\ge10^3$ au) binary
companions is consistent with a population of companions formed through initial
fragmentation \citep{Bonnell:1992ep}. In reality, the
observed population is likely a synthesis of, at least, these two types
of fragmentation. Hydrodynamical simulations which incorporate both of
these processes are able to produce companions over a wide range of
separations \citep{Bate:2009br,Bate:2011hy}, consistent with the
observed separation distribution for solar-type stars, although these
simulations are not yet large enough to include a statistically
significant number of A-type stars.

\subsection{Multiplicity of A-type star subsamples}
Metallic-line (Am) A-type stars are distinct from normal A-type stars
due to an overabundance of heavy elements within their observed
spectra, with the notable exception of calcium and scandium
\citep{Boffin:2010dp}. These anomalous abundances can be explained by diffusion
within the stellar atmosphere \citep{Michaud:1980cz,Talon:2006dp},
which can only occur for stars with rotational velocities of $v \lesssim 100$ km s$^{-1}$ \citep{Michaud:1983fo}. As
they do not posses sufficiently strong magnetic fields to induce
magnetic breaking \citep{Conti:1970fq,Auriere:2010kt}, Am stars must
either form with a lower rotational velocity than normal A-type stars,
or this initial fast rotational velocity is reduced through the tidal
breaking caused by an orbiting companion \citep{Roman:1948ev}. In
order to effectively reduce the rotational velocity of the primary
necessary for diffusion to occur, the companion must be on a relatively short
orbit with $P \lesssim 35$ d \citep{Budaj:1997wm}, consistent with the
observed abundance of short-period spectroscopic companions to Am
stars \citep{Abt:1985bs,Carquillat:2007hj}. For the 50 stars within
the VAST sample classified as metallic-lined within the SIMBAD
data base, an MF of $58.0^{+6.5}_{-7.1}$ per cent was
estimated based on a combination of those companions resolved in this
study, and those previously reported within the literature. This is
significantly higher than the multiplicity fraction for the sample of
371 normal (non-Am non-Ap) A-type stars, at $37.6^{+2.6}_{-2.4}$ per
cent, a discrepancy which can be attributed to the relatively large number of
known spectroscopic components to the Am stars within the VAST sample
(e.g. \citealp{Abt:1962fo,Abt:1985bs,Carquillat:2007hj}).

Like Am stars, chemically peculiar (Ap) A-type stars are also observed to have low
rotational velocities \citep{Abt:1972hm}. The low frequency of
close binary companions to Ap stars is inconsistent with the tidal breaking
mechanism used to explain the low rotational velocities of Am stars
\citep{Babcock:1958fz,Carrier:2002hs}, and instead an
alternate mechanism involving magnetic breaking
is thought to predominate \citep{Abt:1973dd}. The strong magnetic
fields measured in non-HgMn Ap stars precludes the presence of a close binary
companion \citep{Abt:1973dd}, consistent with the observed deficiency of binary
companions with periods $\lesssim 3$ days \citep{Carrier:2002hs}. With
the full VAST sample, only 14 stars are classified as being chemically
peculiar (Ap) within the SIMBAD data base. Combining those companions
measured within this study with known literature companions (Table
\ref{tab:literature_binaries}), we find an MF
of $50.0^{+12.4}_{-12.4}$ per cent for this Ap subsample, somewhat
larger than, but still consistent with, the MF of
the non-Am non-Ap subsample of 371 stars of $37.6^{+2.6}_{-2.4}$ per cent.

Another subsample of A-type stars of particular note are those known
to host extrasolar planets, of which two such examples exist within the
VAST sample - $\beta$ Pic (HIP27321; \citealp{Lagrange:2009hq}) and
HR 8799 (HIP 114189; \citealp{Marois:2008ei,Marois:2010gp}). No
stellar companions were identified to either of these stars within the
AO observations presented within this study, and no wide
CPM companions were identified up to a separation of 45,000 au around either HR
8799, consistent with previously published analyses
\citep{Close:2009fu} and $\beta$ Pic. Although $\beta$ Pic was not
originally included in the wide CPM companion search described in
Section 4.2 due to its relatively low proper motion, it is included
here for completeness. This subsample was expanded by including all
known exoplanet-hosting main sequence A-type stars which were not
included within the VAST survey - Fomalhaut \citep{Kalas:2008cs},
$\kappa$ And \citep{Carson:2013fw} and HD 95086
\citep{Rameau:2013dr}. Excluding the known comoving companions to
Fomalhaut, TW PsA \citep{Luyten:1938fs,Shaya:2010dt} and LP
  876-10 \citep{Mamajek:2013tr}, no additional
wide CPM companions up to a separation of 45,000 au were identified using the
same procedure as described in Section 4.2. Assuming that all of the
stellar components within each system have been resolved, the
MF of the known exoplanet-hosting A-type stars can
be estimated as 20$^{+25}_{-8}$ per cent, a preliminary approximation
given the extremely small sample size.

\subsection{Higher order multiples}
\begin{figure*}
\resizebox{\hsize}{!}{{\includegraphics{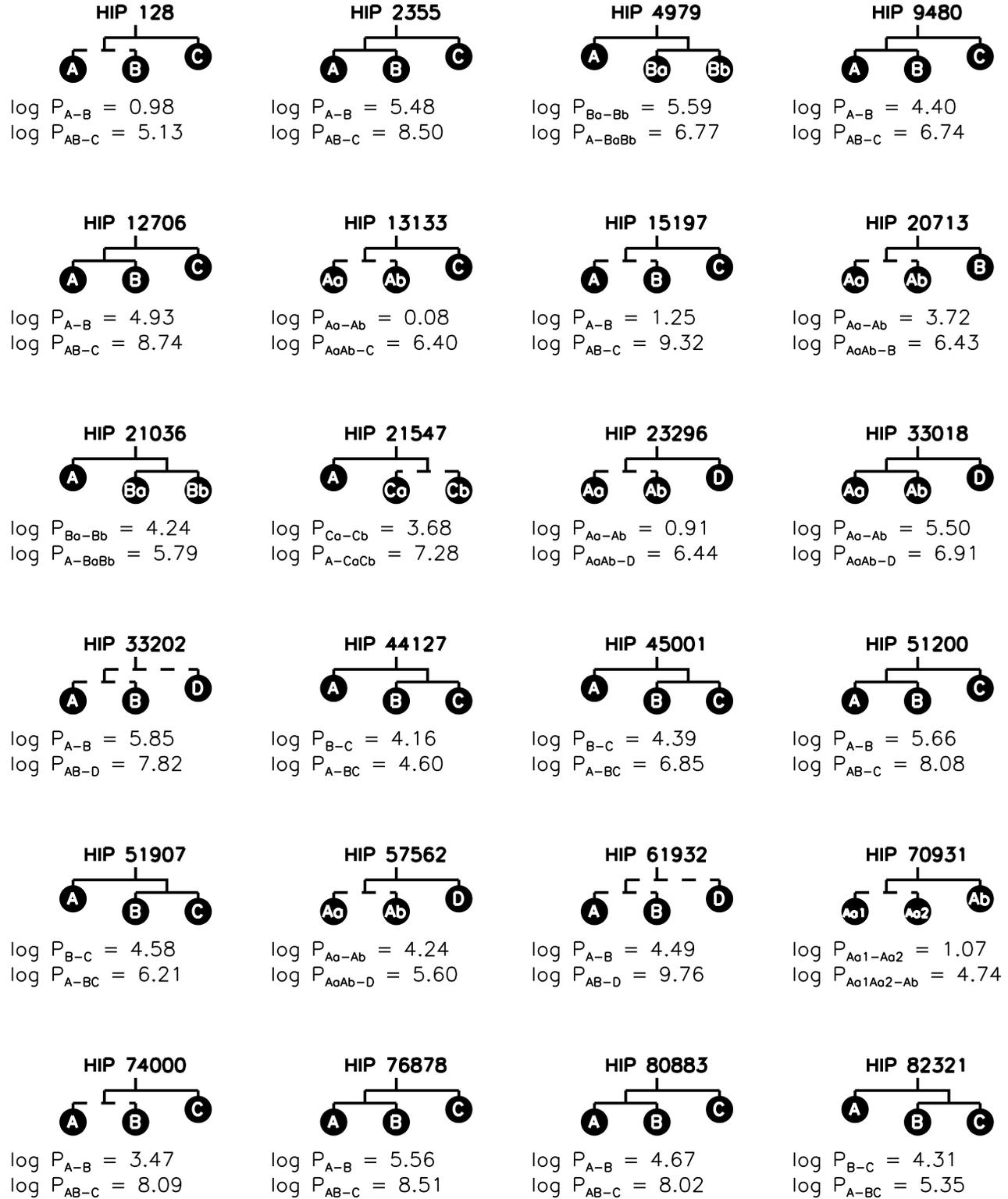}}} 
\caption{Schematic representations of the higher order multiple
  systems, combining those systems resolved within this study and
  previously known systems obtained from the literature. The
  components which were not resolved within this study are indicated
  by a dashed connection - for example, the B component of the HIP
  128 system was not resolved as it is a spectroscopic binary with a
  $\sim$10 day period. The period is given for each hierarchical
  system, obtained from either a spectroscopic orbit fit, or an
  estimate based on the projected physical separation.}
\label{fig:mobile}
\end{figure*}
\addtocounter{figure}{-1}
\begin{figure*}
\resizebox{\hsize}{!}{{\includegraphics{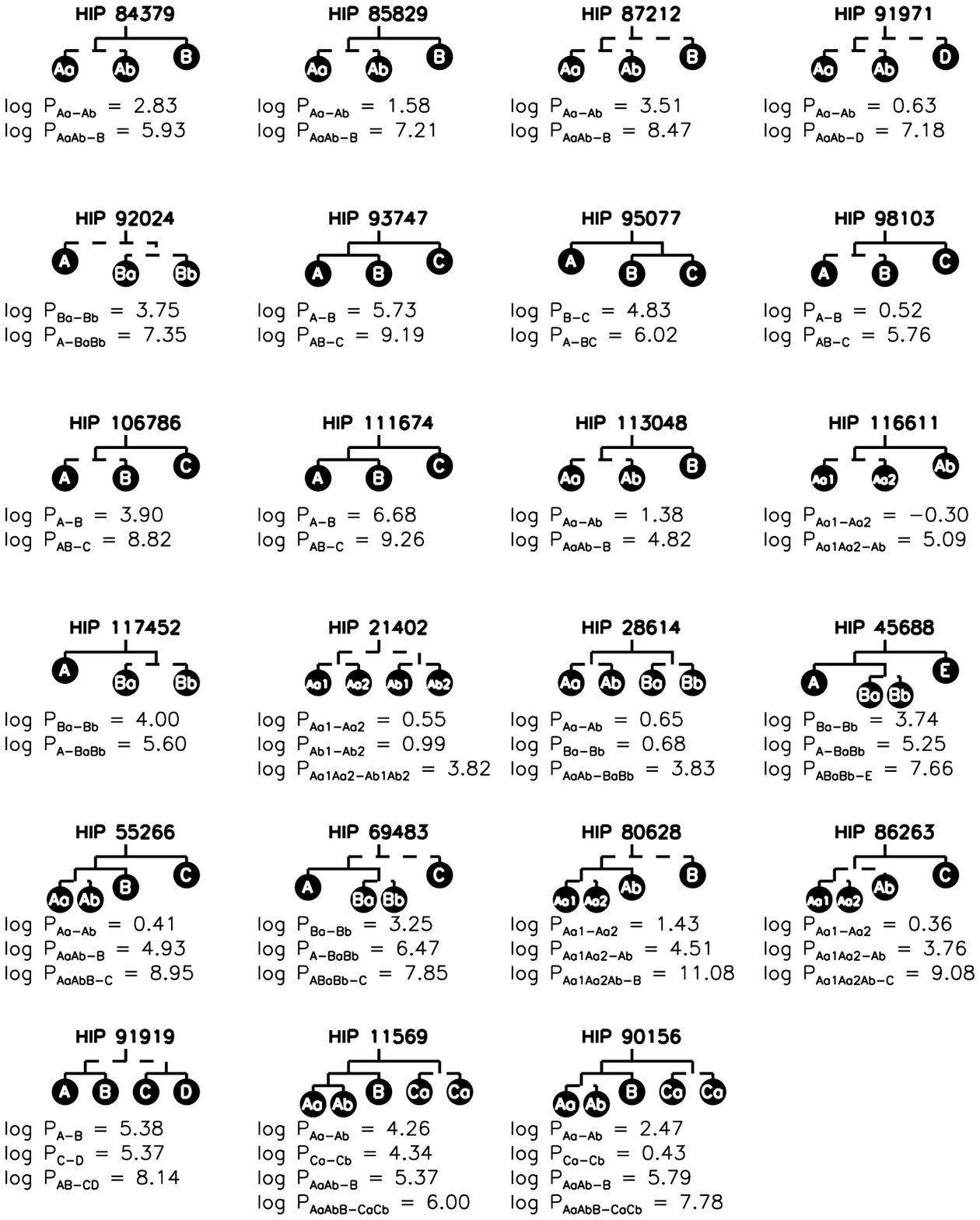}}} 
\caption{{\it(Continued)} Schematic representations of the higher order multiple
  systems, combining those systems resolved within this study and
  previously known systems obtained from the literature. The
  components which were not resolved within this study are indicated
  by a dashed connection - for example, the B component of the HIP
  128 system was not resolved as it is a spectroscopic binary with a
  $\sim$10 day period. The period is given for each hierarchical
  system, obtained from either a spectroscopic orbit fit, or an
  estimate based on the projected physical separation.}
\end{figure*}
\begin{figure}
\resizebox{\hsize}{!}{{\includegraphics{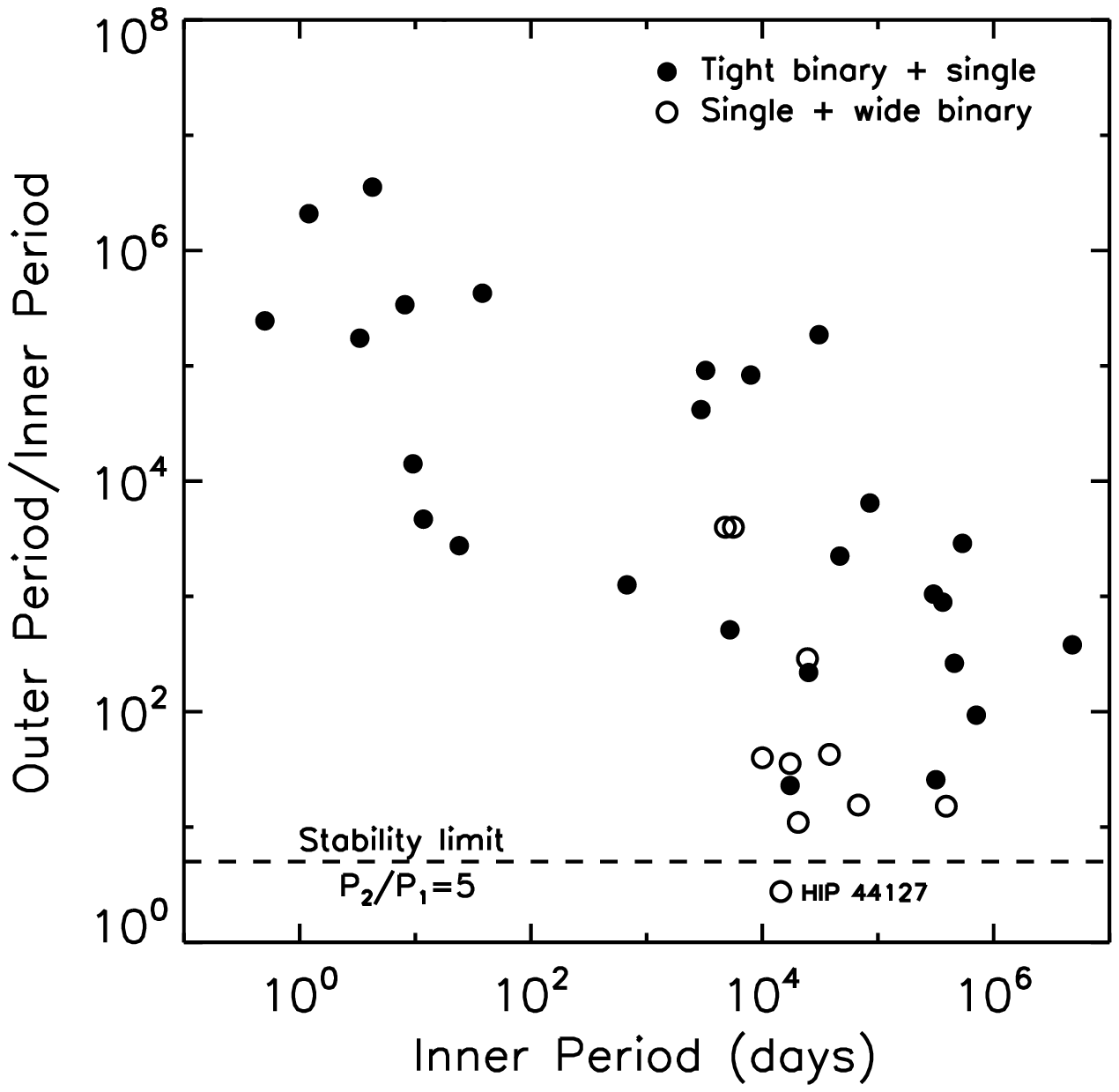}}} 
\caption{The ratio of the outer and inner period plotted as a function
  of the inner period for hierarchical triple systems consisting of a
 tight binary and a wide tertiary component (filled circles), and those consisting of a
  single star with a distant pair of lower mass components in a tight
  orbit (open circles). For binary systems without an orbital period,
  the period has been estimated from the projected separation. The
  stability limit of $P_2/P_1   =5$ is also shown (dashed
  line), with systems below this line being susceptible to dynamical processing
  \citep{Eggleton:2006wx}. The only hierarchical triple system found
  to be in a potentially unstable configuration within this study is indicated.}
\label{fig:hierarchical}
\end{figure}
Separations of less than 30 au are not fully covered by the VAST survey, but
combining the spatially resolved systems from this study, with
spectroscopic, speckle and interferometric binaries
\citep{Mason:2001ff,Pourbaix:2004dg}, allowed for a
lower limit estimate of the population of higher order
multiples. Among the 156 VAST targets which had both AO observations
and were searched for CPM companions - providing sensitivity to the widest
range of separations - the number of single, binary, triple, quadruple and quintuple
systems were 88, 50, 14, 3 and 1, respectively ($56.4_{-4.0}^{+3.8}$,
$32.1_{-3.5}^{+3.9}$, $9.0_{-1.8}^{+2.8}$, $1.9_{-0.6}^{+1.8}$, $0.6_{-0.2}^{+1.4}$
per cent). These relative frequencies are comparable to the ratio of
multiple systems measured for solar-type stars
\citep{Raghavan:2010gd}, although the A-type star frequencies will
change as new components are resolved. Expanding the analysis to
include the full VAST sample of 435 stars, the relative frequencies
remain statistically consistent, with the number of single, binary,
triple, quadruple and quintuple systems found to be 259, 129, 37, 8
and 2, respectively ($59.5_{-2.4}^{+2.3}$,
$29.7_{-2.1}^{+2.3}$, $8.5_{-1.1}^{+1.5}$, $1.8_{-0.4}^{+0.9}$,
$0.5_{-0.1}^{+0.6}$ per cent). Schematic representations of the 47
systems with three or more components are given in Fig. \ref{fig:mobile}.

Of the 37 triple systems, 27 consist
of a tight binary system orbited by a distant, typically lower mass
component. The remaining 10 systems are composed of a single primary,
with a distant pair of lower mass components in a tight orbit. A
preference of hierarchical triple systems consisting of a tight binary
orbited by a more distant tertiary is found for solar-type
primaries \citep{Raghavan:2010gd}, and while a similar preference is
observed for A-type primaries, significant observational biases still
exist. The greater frequency of this type of system would suggest that
a more distant pair in a wide orbit around a single primary is either
formed less frequently, or more susceptible to dynamical processing. Fig.
\ref{fig:hierarchical} shows the ratio of the inner and outer period
plotted as a function of the inner period for the 37 triple
systems. All of the triple systems with a single primary and a distant
pair in a tight orbit are found closer to the stability limit; however, 
this also coincides with the portion of the diagram to which the AO
observations are sensitive to companions. To determine if the clustering of these
single plus wide binary systems near the stability limit is simply an
observational bias, high-resolution measurements of the secondary
components of wide binary systems identified within this study are
required in an attempt to resolve them into hierarchical triples.

Of the eight known quadruple systems, three are of $\epsilon$ Lyr-type,
consisting of two pairs of binaries, with similar mass ratios and
orbital periods, in a wide orbit \citep{Tokovinin:2008bl}. The
remaining five quadruple systems, and the two higher order systems,
are all multi-levelled hierarchical systems,
potentially products of multiple fragmentation events during the star
formation process. All of these hierarchical multiple systems
represent ideal candidates for orbital motion monitoring work. With
sufficient orbital coverage, model-independent mass estimates can be
derived for each component within the system
(e.g \citealp{Kohler:2012gc}), a useful diagnostic of evolutionary models
(e.g. \citealp{DeRosa:2012gq}). 

Two of the hierarchical systems within this study were found to have
low period ratios, indicative of a dynamically unstable system
\citep{Eggleton:2006wx}; a triple system consisting of a
single primary and a wide pair in a tight orbit (HIP 44127 A-BC;
$\iota$ UMa), and the widest binary component  of a higher order system (HIP 11569 AB-CaCb; $\iota$ Cas). Both of these
systems have indications of youth
\citep{Plavchan:2009fv,DeRosa:2012gq}, which would be consistent 
with an unstable dynamical configuration
\citep{Zhuchkov:2012ey}. As the outer period in both cases
are based on the projected separation measured within the AO
observations, the systems cannot be conclusively described as unstable
-- measurements of the true semi-major axes are required.

\section{Summary}
We have obtained high angular resolution AO observations of 363
nearby ($D \le 75$ pc) A-type stars in order to characterize the
population of binary companions to these intermediate mass
($M=1.5-3.0$ M$_{\odot}$ stars. Combining these AO data with a search
for wide CPM companions of 228 A-type stars using all-sky photographic
plates, resulted in sensitivity to separation ranges of 30 -- 800 au
for the AO data and 1,800 -- 45,000 au for the CPM search. A total of
137 companions were identified, consisting of 113 AO companion
candidates with $\ge 95$ per cent chance of being physically
associated and 24 wide CPM companions confirmed by an analysis of
their proper motion and position on the CMD. Of the resolved binary
population, 64 were newly identified as companions as a result of this
study.

Over a restricted separation range of 30 -- 800 au, we measure a
CSF of $21.9 \pm 2.6$ per cent (Fig.
\ref{fig:multiplicity}), compared with $19.6 \pm 2.1$ per cent for
solar-type primaries \citep{Raghavan:2010gd}, and $17.1 \pm 5.4$ per
cent for M-dwarf primaries \citep{Fischer:1992cj}. The results are
indicative of a trend of increasing multiplicity as a function of
increasing primary mass, although the significant error bars prevent
this from being a conclusive result (Fig. \ref{fig:multiplicity}
and \ref{fig:multiplicity_fit}). This trend is consistent with
predictions of the frequency of binary systems determined from
numerical simulations
\citep{Sterzik:2003jh,Hubber:2005it,Bate:2009br,Bate:2011hy}, although
the functional form of the increase does differ in all but the latter
simulation (Fig. \ref{fig:multiplicity_simulations}).

The distribution of companion mass ratios over the 30 -- 800 au range
was found to consist of two statistically distinct distributions,
consisting of AO companions resolved interior to, and exterior to, a
separation of $\sim$125 au. The inner distribution was measured to be
flat for companions with $q\ge 0.15$, while wider systems
consisted preferentially of lower mass companions (Fig.
\ref{fig:2qdist}). By performing a similar analysis on the field
solar-type data \citep{Raghavan:2010gd}, we find a similar pattern
of a greater frequency of lower mass companions at wider separations
(Fig. \ref{fig:gqdist}). The relative abundance of equal-mass
companions at close separations in both this study, and for solar-type
primaries, is consistent with the equalisations of the masses of companions at close separations
\citep{Bate:1997uf,Bate:2000ir}. For wide companions ($a_{\rm proj}
\gtrsim 125$ au), the high frequency of lower mass companions may be
indicative of a large population of companions formed through either
initial \citep{Bonnell:1992ep} or disc fragmentation \citep{Kratter:2009gs}.

We have presented the first comprehensive multiplicity statistics for
A-type stars over a wide separation range, with sensitivity extending
to the lowest mass stellar companions. These results are crucial for
providing empirical comparisons to theoretical predictions of binary
star formation over a range of stellar masses, and companion
separations, and comparisons to previous multiplicity surveys of
lower mass primaries. An important extension of this survey would be
to search for companions interior to the detection limits of the AO
data presented within this paper ($a_{\rm proj} \lesssim 30$
au). Interferometric and spectroscopic monitoring of these targets
will provide sensitivity to companions interior to this limit, and
future high-contrast AO instruments can be used to tightly constrain
the population of low-mass stellar companions to these stars within a
more restricted separation range of 10 -- 100 au.

\section*{Note added in proof}
While the manuscript was being proofed,
an investigation of the companion to HIP 96313 resolved within the AO
data set revealed it to be a background giant star, based on its
brightness and extremely red colour. At $\sim$930 au, it lies outside of
the projected separation ranges considered when constructing the
separation and mass ratio distributions (30--800 au and 1,800--45,000 au), and
as such this misidentification has a minimal effect on the final
statistics presented within this study. An investigation performed on
the remaining AO-resolved binary companions revealed no additional
contaminations of a similar nature.

\section*{Acknowledgements}
The authors wish to express their gratitude for the constructive
comments received from the referee, H. A. Abt. The authors wish to
thank M. R. Bate, R. J. Parker and K. M. Kratter for their useful comments
and suggestions which helped to significantly improve the paper. The 
authors also wish to thank N. J. McConnell for helping to obtain a subset of the AO
observations, and his contributions to various aspects of
this study. The authors gratefully acknowledge
several sources of funding. RJDR was funded through a studentship
from the Science and Technology Facilities Council (STFC)
(ST/F007124/1). RJDR gratefully acknowledge financial support
received from the Royal Astronomical Society to fund collaborative
visits. JP is funded through support from the Leverhulme
Trust (F/00144/BJ) and the STFC (ST/F003277/1,
ST/H002707/1). AV acknowledges support from the STFC grant
ST/H002707/1. Portions of this work were performed under the auspices
of the U.S. Department of Energy by Lawrence Livermore National
Laboratory in part under Contract W-7405-Eng-48 and in part under
Contract DE-AC52-07NA27344, and also supported in part by the NSF
Science and Technology CfAO, managed by the UC Santa Cruz under
cooperative agreement AST 98-76783. This work was supported, through
JRG, in part by University of California Lab Research Programme
09-LR-118057-GRAJ and NSF grant AST-0909188. Based on observations 
obtained at the Canada-France-Hawaii Telescope (CFHT) which is
operated by the National Research Council of Canada, the Institut
National des Sciences de l'Univers of the Centre National de la
Recherche Scientifique of France, and the University of Hawaii. Based
on observations obtained at the Gemini Observatory, which is operated
by the Association of Universities for Research in Astronomy, Inc.,
under a cooperative agreement  with the NSF on behalf of the Gemini
partnership: the National Science Foundation (United States), the
Science and Technology Facilities Council (United Kingdom), the
National Research Council (Canada), CONICYT (Chile), the Australian
Research Council (Australia),  Minist\'{e}rio da Ci\^{e}ncia e
Tecnologia (Brazil) and Ministerio de Ciencia, Tecnolog\'{i}a e
Innovaci\'{o}n Productiva (Argentina). The authors also wish to extend
their gratitude to the staff at the Palomar Observatory and the
UCO/Lick Observatory for their support and assistance provided during
the course of the observations. This research is partly based on data
obtained from the ESO Science Archive Facility. This research has made use of the
SIMBAD data base, operated at CDS, Strasbourg, France. This publication
makes use of data products from the Two Micron All Sky Survey, which
is a joint project of the University of Massachusetts and the Infrared
Processing and Analysis Center/California Institute of Technology,
funded by the National Aeronautics and Space Administration and the
National Science Foundation. This research has made use of the
Washington Double Star Catalog maintained at the U.S. Naval
Observatory. This research has made use of data obtained from the SuperCOSMOS
Science Archive, prepared and hosted by the Wide Field Astronomy Unit,
Institute for Astronomy, University of Edinburgh, which is funded by
the UK Science and Technology Facilities Council. This research used
the facilities of the Canadian Astronomy Data Centre operated by the
National Research Council of Canada with the support of the Canadian
Space Agency.

\section*{Appendix 1 - Age and Mass Estimates}
In order to convert the measured magnitude difference between an
A-type primary and any resolved companion into a mass ratio, an
estimate of the age of the primary is required due to their rapid
evolution away from the zero-age main sequence. Age estimates can be
obtained for A-type stars by virtue of membership of a known moving
group (e.g. \citealp{Zuckerman:2004ex,daSilva:2009eu}), the detection
and characterization of debris disc excesses
(e.g. \citealp{Rieke:2005hv,Su:2006ce}), or the location
on a colour-magnitude diagram
(e.g. \citealp{Paunzen:1997vj,Song:2001bv}). The limited rotational
breaking experienced by an A-type star during its short lifespan 
precludes age estimation through gyrochronology techniques
\citep{Barnes:2003ga}, and chromospheric indicators are not reliable
due to the breakdown in the correlation between activity and age for
stars with a colour index of $B-V\le0.5$ \citep{Mamajek:2008jz}.

The position of a star on the colour--magnitude diagram is impacted by
several factors -- the presence of a binary companion, rapid rotation,
and metallicity. A binary companion increases the brightness and
shifts the colour to redder values by an amount dependent on the
spectral type of the companion. If the shift in magnitude and colour
induced by the presence of a companion is not removed, the stars age will be
overestimated based on its position on the colour--magnitude
diagram. By incorporating the results of this imaging binary survey
and previous spectroscopic and interferometric results, we have
corrected the positions of the sample stars on the colour--magnitude
diagram in an attempt to eliminate the effect of the companions and to determine
more accurate ages.

Two factors not considered in this study that can influence stellar
ages are stellar rotation and metallicity. Regardless of spin axis
orientation, a rotating star will have an older age estimate than a
non-rotating star \citep{Collins:1985vd}, as the locus of shifted
locations in the colour--magnitude diagram forms a fan-shape extending
to brighter magnitudes for pole-on rotators and redder colours for
equator-on rotators, with the apex at the non-rotating position (see
fig. 1 in \citealp{Collins:1985vd}, and fig. 2 in
\citealp{Gray:2007ws}). While not all A-type stars in this sample have
$v\sin i$ values, the range of measured values is 10 to 317 km
s$^{-1}$, with an average and standard deviation of $124\pm67$ km
s$^{-1}$ based on the 350 with measurements reported in the Extended
Hipparcos Compilation (XHIP; \citealp{Anderson:2011uv}). The effect of
rotation has been assessed to over-estimate ages of early-type stars
by $\sim$$30-50$\%  \citep{Figueras:1998ux}, and becomes prominent
when $v\sin i > 100$ km s$^{-1}$ \citep{Song:2001bv}. The fact that
rotation can only make an A-type star appear older on the colour--
magnitude diagram may be an explanation as to why ages of known moving
group members within the VAST sample, when estimated using the
technique described in this section, were often found to be
significantly older than the canonical moving group age.

The metallicity of a star can bias the age
estimate in both directions --  if a star has a higher metallicity than
assumed for the isochrones, the age will be overestimated, while the 
age will be underestimated if the metallicity is lower than
assumed. Again, not all the stars have measured metallically, but for
the 108 with [Fe/H] values, the average is near solar at
$-0.06\pm0.51$ \citep{Anderson:2011uv}, and as such the net effect on
the overall age distribution should be small. Since the rotation and
metallicity is not measured for all stars, and the theoretical
isochrones used to determine the ages are for the non-rotating case,
we have not adjusted the age estimates for either of these effects.

\begin{table}
\centering
\caption{Literature ages of known moving groups}
\begin{tabular}{cccc}
Moving Group&Number&Age&Age\\
&of Stars&(Myrs)&Reference\\
\hline
\hline
TW Hydrae&1&8&\cite{Stauffer:1995jl}\\
$\beta$ Pictoris&5&12&\cite{Zuckerman:2001go}\\
Tucana-Horologium&8&30&\cite{Zuckerman:2011bo}\\
Argus&2&40&\cite{Torres:2008vq}\\
AB Doradus&4&70&\cite{Zuckerman:2011bo}\\
Castor&2&200&\cite{BarradoyNavascues:1998tg}\\
Ursa Major&9&300&\cite{Zuckerman:2004ex}\\
Hyades&24&625&\cite{Perryman:1998vj}\\
\hline
\end{tabular}
\label{tab:assoc_ages}
\end{table}

\begin{table}
\centering
\caption{Source of literature isochrone ages}
\begin{tabular}{cc}
Reference&Number\\
&of Stars\\
\hline
\hline
\cite{Gerbaldi:1999cv}&4\\
\cite{Janson:2011hu}&1\\
\cite{Laureijs:2002de}&1\\
\cite{Paunzen:1997vj}&3\\
\cite{Rhee:2007ij}&25\\
\cite{Rieke:2005hv}&8\\
\cite{Song:2001bv}&6\\
\cite{Su:2006ce}&1\\
\cite{Tetzlaff:2010gt}&12\\
\cite{Westin:1985un}&3\\
\hline
\end{tabular}
\label{tab:litiso_ages}
\end{table}

\begin{figure}
\resizebox{\hsize}{!}{{\includegraphics{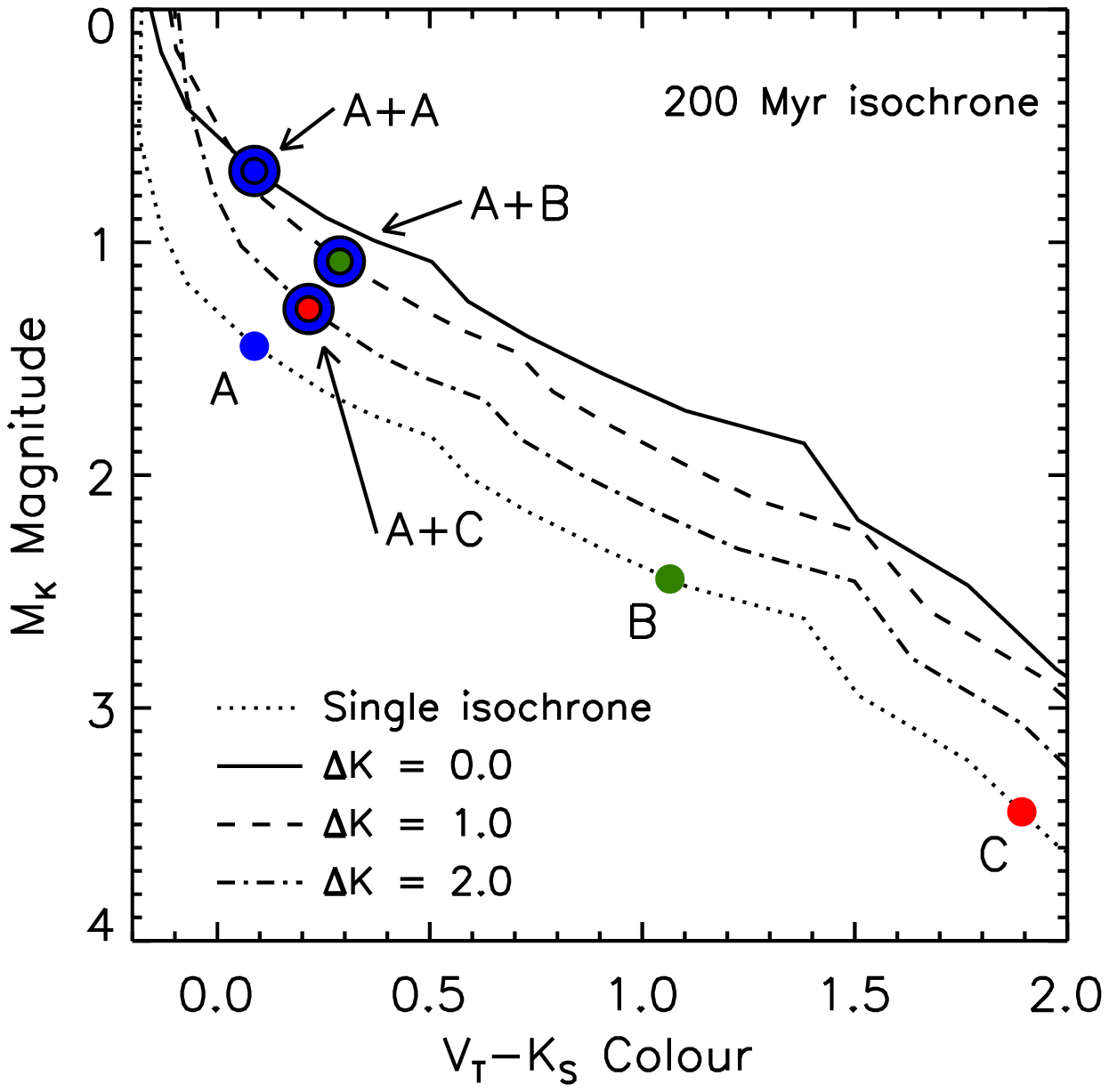}}} 
\caption{The presence of a companion unresolved
within the Tycho2 and 2MASS catalogues can lead
to a significant shift in the position of the primary on the
colour--magnitude diagram. An example is shown demonstrating the
expected shift in the position of a star with an absolute magnitude of
$M_{\rm K}=1.4$ (A), due to the presence of; an equal mass
companion (A+A), a companion with a magnitude difference of $\Delta
K=1$ (A+B) and $\Delta K=2$ (A+C). For the case of a binary system
with two identical components, the primary is only shifted in the
magnitude direction, with no change in the colour of the system. The
presence of a lower mass companion (B, C) can lead to a significant
reddening of the system. This procedure is repeated for each position
within the single star isochrone (dotted line) to produce a binary
isochrone including the effect of a binary companion at range of
magnitude differences ($\Delta K=0$, solid line; $\Delta K=1$, dashed
line; $\Delta K=2$, dot--dashed line).}
\label{fig:binaryiso}
\end{figure}
\begin{figure*}
\resizebox{\hsize}{!}{{\includegraphics{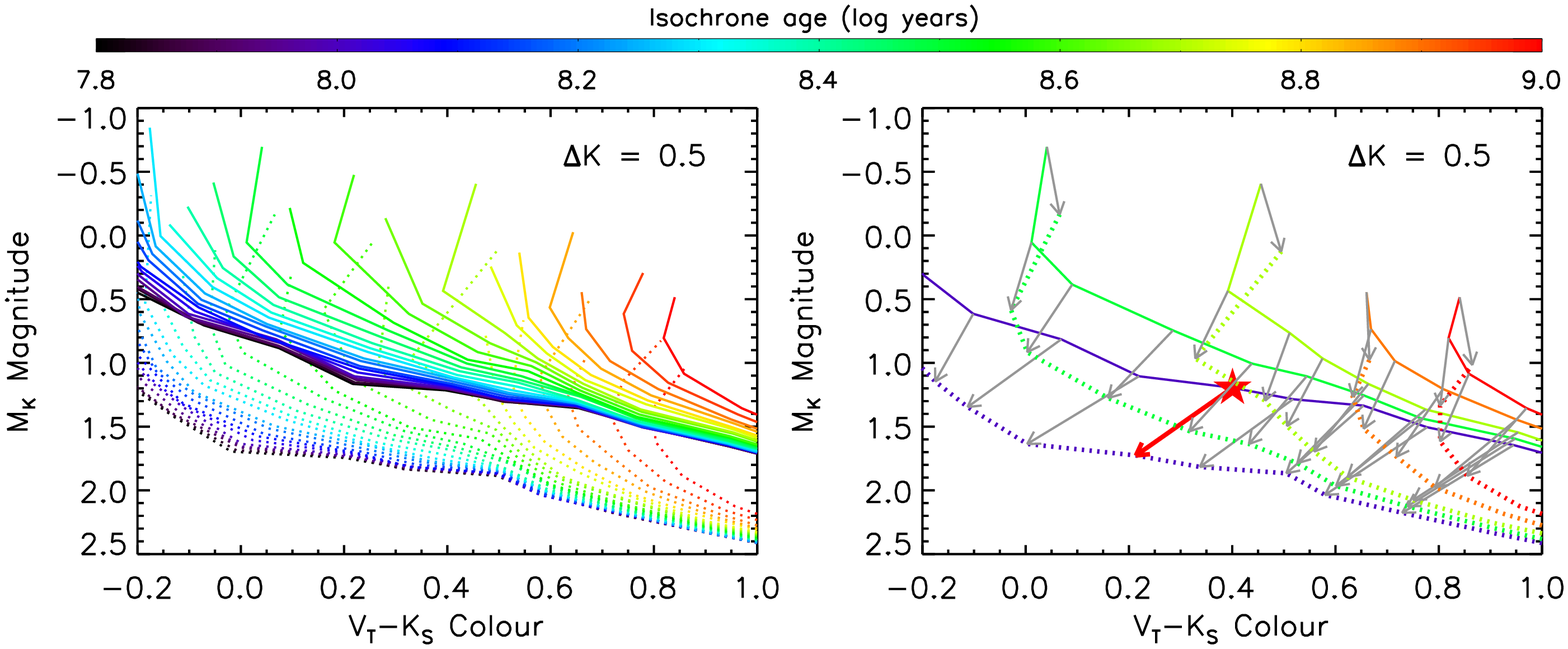}}} 
\caption{Left-hand panel: a set of
binary isochrones was constructed for the full range of isochrone ages
available within the models, for each value of the magnitude
difference. The example shows the single star isochrone (dotted
lines), and the corresponding binary star isochrone for the case of a
$\Delta K=0.5$ companion (solid lines). As the shift in the position
of each point on the isochrone is based on an estimate of the
companion properties derived from the theoretical models, this process
can be carried out for a range of visual magnitude differences
($\Delta V$), for companions resolved with interferometry, or mass
ratios ($q$), for double-lined spectroscopic binaries. Right-hand panel: The magnitude and direction of the shift required to correct
  the position of the primary on the colour--magnitude diagram due to
  the presence of a $\Delta K=0.5$ companion unresolved within the
  Tycho2 and 2MASS catalogues. For clarity only a
  subset of the single star (dotted lines) and binary isochrones
  (solid lines) are shown ($10^8$, $10^{8.5}$, $10^{8.7}$, $10^{8.9}$, and $10^9$ yrs). At each
  point on the binary isochrone, a vector is plotted showing the magnitude and direction of the
  correction required to remove the contamination induced by the
  presence of the companion (grey arrows). As an
  example, for a binary system with an observed $M_{\rm K}(AB)=1.15$ and
  $(V-K)_{\rm AB}=0.4$ (red filled star), with a magnitude difference of $\Delta
  K=0.5$, the  correction estimated for the magnitude and colour is
  shown (red vector). This correction has reduced the age estimate of
  this hypothetical primary from 500 to 100 Myr.}
\label{fig:binaryiso_vectors}
\end{figure*}
\begin{figure}
\resizebox{\hsize}{!}{{\includegraphics{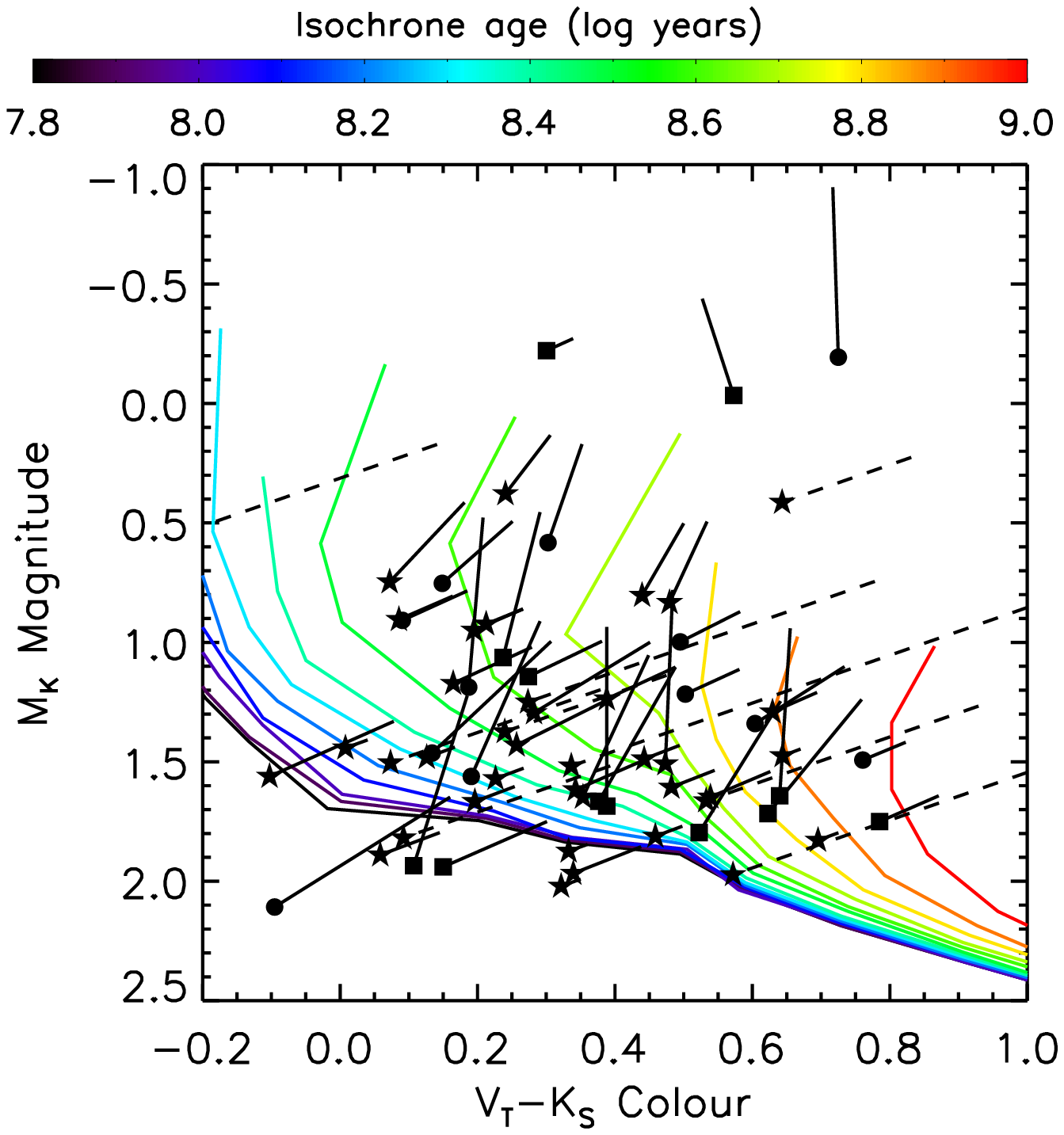}}} 
\caption{The position of 61 stars is changed due to a
  presence of a bright binary companion unresolved in either one of,
  or both, the Tycho2 and 2MASS catalogues. The
  magnitude and direction of the shift necessary to correct the star
  (solid line - unresolved in both catalogues, dashed line -
  unresolved only in 2MASS), is estimated based on the predicted colour and
  magnitude of the companion derived from the theoretical
  isochrones. The corrected position of each star is given by a filled
  symbol, depending on the method through which the companion was
  resolved. The companion properties are estimated using either the
  magnitude difference measured within this study ($\Delta K$, filled stars), the
  magnitude difference reported for companions resolved with
  interferometric techniques ($\Delta V$, filled circles), or the mass ratio
  calculated for a double-lined spectroscopic binary ($q$, filled
  squares). Theoretical isochrones from \citet{Siess:2000tk} are
  plotted within the range $7.8 \le \log t \le 9.0$, in steps of $\Delta \log t
  = 0.1$.}
\label{fig:appendix_cmdcorrection}
\end{figure}
\begin{figure}
\resizebox{\hsize}{!}{{\includegraphics{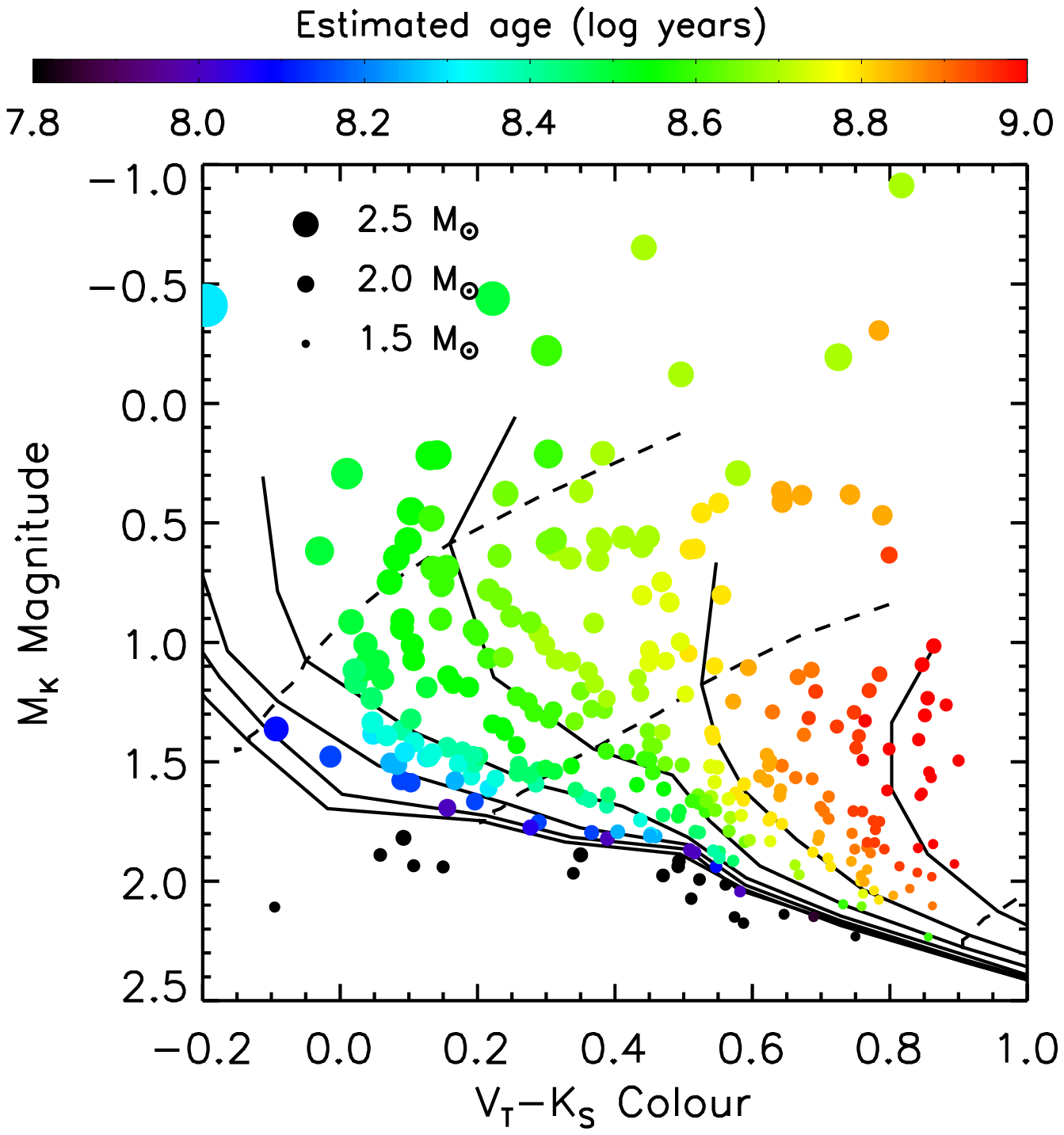}}} 
\caption{A colour--magnitude diagram of the 316 stars
  without age estimates presented within the literature. The colour of each symbol denotes the age estimated
  for each star based on a comparison to theoretical solar-metallicity
isochrones \citep{Siess:2000tk}, with the size being proportional to
the mass estimated from the theoretical mass-magnitude
relations. The theoretical isochrones are plotted within the range $7.8 \le
\log t \le 9.0$, in steps of $\Delta \log t = 0.2$ (solid lines),
alongside the theoretical evolutionary tracks for a  2.5, 2.0 and 1.5
M$_{\odot}$ star (dashed lines, left to right).}
\label{fig:appendix_cmdsingle}
\end{figure}

Of the 435 stars with either AO
observations, or with proper motions sufficient for the search for
CPM companions, 55 are known members of
coeval moving groups or stellar clusters with literature ages
(Table \ref{tab:assoc_ages}), and 64 have age estimates within the
literature derived from theoretical evolutionary models (Table
\ref{tab:litiso_ages}). For the remaining 316 stars, the age was
estimated based on a comparison of the position of the star on the $V-K$ versus
$M_K$ colour--magnitude diagram against theoretical stellar isochrones
\citep{Siess:2000tk}. As the presence of a bright binary companion, at
an angular separation less than the resolving limit of the
Tycho2 and 2MASS catalogues, can introduce a
significant shift in the position of a star on the colour--magnitude
diagram, the colour and magnitude of the primary was corrected based
on the expected $V-K $ colour and $M_{K}$ magnitude of the
companion. The magnitude and direction of these corrections were estimated
by constructing a series of isochrones which are shifted due to the
presence of a hypothetical binary companion (Fig.
\ref{fig:binaryiso}). Given that the magnitude difference between the
primary and companion is known, or the mass ratio of the system in the
case of double-lined spectroscopic binaries, the set of binary
isochrones corresponding to the configuration of the system can be
selected, and used to obtain the predicted shift in the position of
the primary on the colour--magnitude diagram (Fig.
\ref{fig:binaryiso_vectors}). Combining the companions resolved within
this study, with binary companions reported within either the WDS or
SB9 catalogues, the positions of 61 stars were corrected (Fig.
\ref{fig:appendix_cmdcorrection}).

\begin{figure*}
\resizebox{\hsize}{!}{{\includegraphics{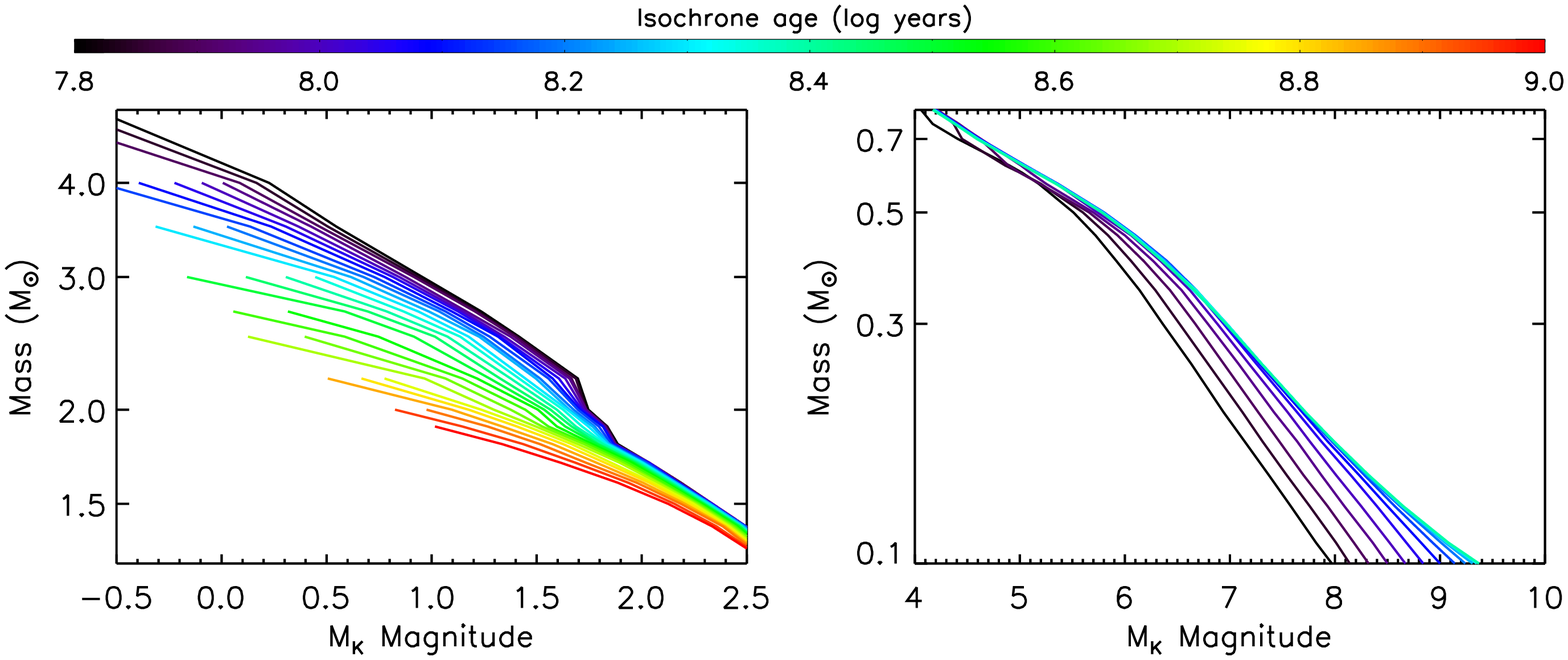}}} 
\caption{The mass-magnitude relations derived from the theoretical
  isochrones covering the A-type star mass range
  (\citealp{Siess:2000tk}; left-hand panel), and for companions at the
  bottom of the Main Sequence (\citealp{Baraffe:1998ux}; right-hand
  panel). For A-type stars, the mass derived from the mass-magnitude
  relation is strongly dependent on the age estimate of the system
  over the full lifespan of a typical A-type star,  while for
  lower mass companions this dependency rapidly becomes negligible
  at ages older than $10^8$ yrs. This demonstrates why an age estimate
is necessary prior to estimating the mass of the A-type star, and mass
ratio of any resolved binary systems.}
\label{fig:massmag}
\end{figure*}

With an age estimates obtained for the observed sample (Fig. \ref{fig:age}, Table
\ref{tab:sample}), a mass-magnitude relation was constructed for each target in
order to estimate the mass of the star and any resolved companions
(Fig. \ref{fig:massmag}, left-hand panel). The increasing luminosity of
an A-type star, as it evolves away from the zero-age main sequence,
leads to a dependence on the mass derived from mass-magnitude
relations as a function of age. Similarly, for the youngest M-dwarf
companions resolved within this study, a dependence on age of the
mass-magnitude relation is caused by the contraction of these
low-mass stars on to the main sequence within the first 100 Myr of
their lifespan, significantly decreasing their luminosity
(e.g. \citealp{Stauffer:1980jz}; Fig. \ref{fig:massmag}, right-hand
panel). For each star within the sample, the mass is estimated using
the $K$-band magnitude obtained from the 2MASS catalogue,
after correction due to the presence of any known companions within the
resolution limit of the 2MASS observations (Fig.
\ref{fig:appendix_cmdcorrection}). The distribution
of estimated masses is given in Fig. \ref{fig:mass}, with the
estimate for each star presented in Table \ref{tab:sample}.
\FloatBarrier

\bibliographystyle{mn2e}
\bibliography{paper}
\setcounter{table}{0}
\onecolumn
\begin{landscape}
{\footnotesize
\begin{longtable}{ccr@{$\pm$}lr@{$\pm$}lr@{$\pm$}lr@{$\pm$}lr@{$\pm$}lr@{$\pm$}lcccr@{$~\le \log a <~$}lr@{$~\le \log a <~$}l}
\caption{The VAST sample}\\
HIP&Spectral&\multicolumn{2}{c}{Distance}&\multicolumn{2}{c}{$B_{\rm
    T}$}&\multicolumn{2}{c}{$V_{\rm
    T}$}&\multicolumn{2}{c}{$J$}&\multicolumn{2}{c}{$H$}&\multicolumn{2}{c}{$K_{\rm
    S}$}&Age$^{a}$&Age&Mass&\multicolumn{4}{c}{Separation
  coverage ($\log a$ [au])}\\
&type&\multicolumn{2}{c}{\textit{(pc)}}&\multicolumn{2}{c}{\textit{(mag)}}&\multicolumn{2}{c}{\textit{(mag)}}&\multicolumn{2}{c}{\textit{(mag)}}&\multicolumn{2}{c}{\textit{(mag)}}&\multicolumn{2}{c}{\textit{(mag)}}&\textit{(Myrs)}&Ref.&\textit{($M_{\odot}$)}&\multicolumn{2}{c}{Adaptive
  optics}&\multicolumn{2}{c}{CPM search}\\
\hline
\hline
\endhead       
128&Am...&70.8&1.7&6.73&0.01&6.52&0.01&6.10&0.02&6.06&0.04&6.02&0.02&180&1&1.84&1.50&2.90&3.60&4.65\\
159&A3&62.5&2.1&7.26&0.01&6.96&0.01&6.33&0.02&6.27&0.02&6.21&0.02&60&1&1.59&1.50&2.90&\multicolumn{2}{c}{--}\\
1473&A2V&41.3&0.4&4.59&0.01&4.51&0.01&4.34&0.27&4.42&0.18&4.46&0.29&200&1&2.26&1.50&2.55&3.60&4.65\\
2355&A7III&62.8&1.7&5.52&0.01&5.25&0.01&4.86&0.25&4.69&0.19&4.46&0.03&710&1&2.20&1.50&2.90&3.60&4.65\\
2381&A3V&53.1&0.8&5.32&0.01&5.18&0.01&5.28&0.24&4.88&0.08&4.83&0.02&450&1&2.09&1.50&2.90&\multicolumn{2}{c}{--}\\
2472&A0V&53.0&0.7&4.79&0.01&4.75&0.01&4.67&0.04&4.77&0.07&4.70&0.03&320&1&2.37&1.50&2.90&\multicolumn{2}{c}{--}\\
2578&A0V&45.6&0.4&5.12&0.01&5.07&0.01&5.06&0.04&5.16&0.08&4.99&0.02&30&2&2.23&1.50&2.90&\multicolumn{2}{c}{--}\\
2852&A5m...&48.9&0.8&6.39&0.01&6.09&0.01&5.56&0.02&5.51&0.04&5.42&0.02&500&1&1.66&1.50&2.55&\multicolumn{2}{c}{--}\\
\hline
\end{longtable}}
\end{landscape}
{\footnotesize
\begin{longtable}{cccccccc}
\caption{Alternative catalogue identifiers}\\
HIP&Name&Bayer&Flamsteed&HR&HD&ADS&WDS\\
\hline
\hline
\endhead
128&&&&&224890&&\\
159&&&&&224945&&\\
1473&&$\sigma$ And&25 And&68&1404&&\\
2355&&&28 And&114&2628&409 AB&J00301+2945\\
2381&&&&118&2696&&\\
2472&&$\lambda^1$ Phe&&125&2834&&J00314-4848\\
2578&&$\beta^3$ Tuc&&136&3003&&J00327-6302\\
2852&&&&151&3326&&\\
\hline
\end{longtable}}
\label{lastpage}
\end{document}